\documentclass[11pt]{article}

 \usepackage{empheq}

 \usepackage{hyperref}
 \usepackage{dsfont}
\usepackage{setspace}
\usepackage{float}
\floatstyle{plaintop}
\restylefloat{table}
\usepackage{tablefootnote}
\usepackage{placeins}
\usepackage{amsmath, hyperref, wasysym}
\usepackage{latexsym}
\usepackage{cite}
\usepackage{lscape}
\usepackage{filecontents}
\usepackage{amssymb}
\usepackage{lineno}
\usepackage[margin=0.615in]{geometry}
\usepackage{wasysym}
\usepackage{fancyhdr}
\usepackage{graphicx}
\usepackage{times}
\usepackage{color,soul}
\usepackage{cancel}
\usepackage{multicol}
\usepackage{placeins}
\usepackage{authblk}
\usepackage{arydshln}
\usepackage{float}
\usepackage{dsfont}
\usepackage{caption}
\usepackage{subcaption}
\usepackage{booktabs}
\usepackage{enumerate}
\usepackage{enumitem}
\usepackage{adjustbox}
\restylefloat{table}
\usepackage{lipsum}
\usepackage[labelfont=bf]{caption}

\makeatletter
 \newcommand\gleftrightarrow[2][]{
   \ext@arrow 9999{\longleftrightarrowfill@}{#1}{#2}}
 \newcommand\longleftrightarrowfill@{
   \arrowfill@\leftarrow\relbar\rightarrow}
  \newcommand{\pfeffer}[0]{\textit{pfeffer}}
    \newcommand{\shady}[0]{\textit{shady}}
      \newcommand{\nacre}[0]{\textit{nacre}}
\makeatother

\begin{document}

\title{Topological data analysis of zebrafish patterns}
\author[a, $\ast$]{Melissa R. McGuirl}
\author[b]{Alexandria Volkening}
\author[a, c]{Bj\"{o}rn Sandstede}

\affil[a]{Division of Applied Mathematics, Brown University, Providence, RI 02912}
\affil[b]{NSF-Simons Center for Quantitative Biology, Northwestern University, Evanston, IL 60208}
\affil[c]{Data Science Initiative, Brown University, Providence, RI 02912}
\affil[$\ast$]{Email address for correspondence: melissa\_mcguirl@brown.edu}

\maketitle

\begin{abstract}
Self-organized pattern behavior is ubiquitous throughout nature, from fish schooling to collective cell dynamics during organism development. Qualitatively these patterns display impressive consistency, yet variability inevitably exists within pattern-forming systems on both microscopic and macroscopic scales. Quantifying variability and measuring pattern features can inform the underlying agent interactions and 
allow  for predictive analyses. Nevertheless, current methods for analyzing patterns that arise from collective behavior only capture macroscopic features, or rely on either manual inspection or smoothing algorithms that lose the underlying agent-based nature of the data. Here we introduce methods based on topological data analysis and interpretable machine learning for quantifying both agent-level features and global pattern attributes on a large scale. Because the zebrafish is a model organism for skin pattern formation, we focus specifically on analyzing its skin patterns as a means of illustrating our approach. Using a recent agent-based model, we simulate thousands of wild-type and mutant zebrafish patterns and apply our methodology to better understand pattern variability in zebrafish. Our methodology is able to quantify the differential impact of stochasticity in cell interactions on wild-type and mutant patterns, and we use our methods
to predict stripe and spot statistics as a function of varying cellular communication. Our work provides a new  approach to automatically quantifying biological patterns and analyzing agent-based dynamics so that we can now answer critical questions in pattern formation at a much larger scale.
\end{abstract}


\section*{Introduction}

\begin{figure*}[ht!]
\centering
\includegraphics[width=\textwidth]{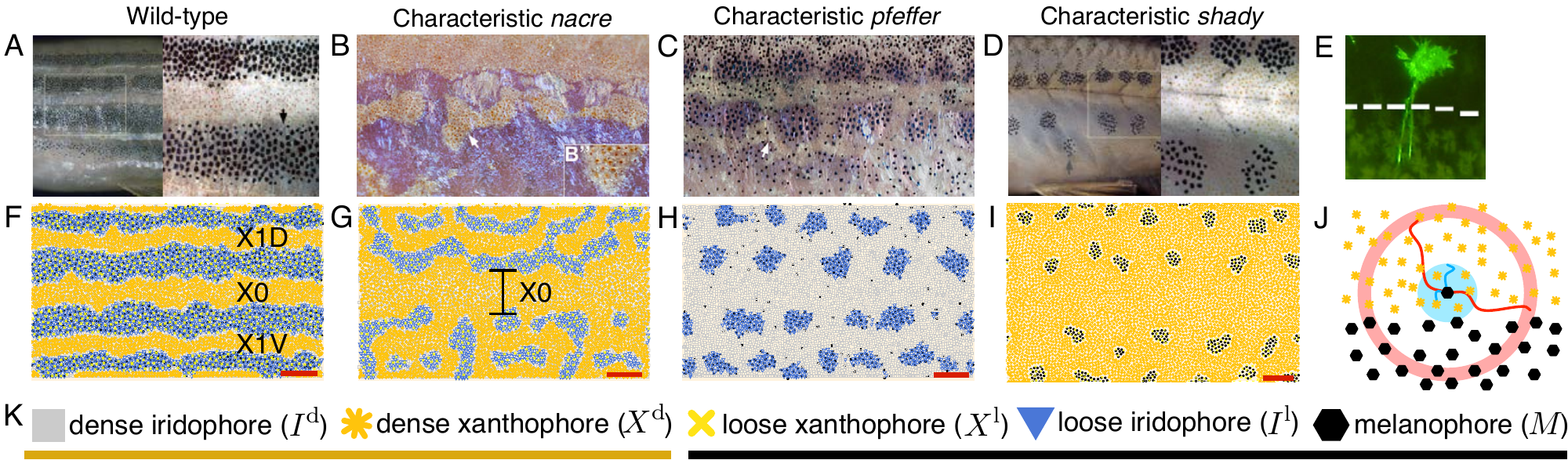}
\caption{\label{motivation} Self-organization during development: diverse skin patterns form on zebrafish due to the interactions of pigment cells. (A) Wild-type zebrafish feature dark stripes and light interstripes \cite{Jan,Frohnhofer}, while mutant patterns that form because a particular cell type is missing have altered, more variable patterns. (B) \textit{Nacre} (encoding mitfa) \cite{Maderspacher2003,Lister} has an enlarged central orange region flanked by blue patches. (C) \textit{Pfeffer} (encoding csf1rA) \cite{PatDev127,ParTur130,Maderspacher2003} is characterized by messy spots arranged horizontally \cite{Frohnhofer}. (D) The \textit{shady} mutant (encoding ltk) \cite{Frohnhofer,Lopes} often features smooth black spots roughly arranged in stripes. (E) During pattern formation, pigment cells extend long legs (measuring up to half a stripe width in distance) toward interstripe cells for communication \cite{delta}. (F--I) The agent-based model \cite{volkening2} replicates zebrafish patterns \emph{in silico}. (The central light interstripe is often labeled \emph{X0}, and the next two interstripes are called \emph{X1V} and \emph{X1D} \cite{Frohnhofer}.) (J) Rules for agent behavior in the model \cite{volkening2} depend on the cells in short-range disks and a long-range annulus. (K) Summary of the main pigment cells involved in patterning: interstripes consist of orange dense xanthophores and silver dense iridophores, and dark stripes contain yellow loose xanthophores, blue loose iridophores, and black melanophores. Images (A--D) reproduced from Frohnh\"{o}fer \emph{et al}.\ \cite{Frohnhofer} and image (E) reproduced from Hamada \emph{et al}. \cite{delta} with adaption; images (A--E) licensed under CC-BY 3.0 (http://creativecommons.org/licenses/by/3.0) and published by The Company of Biologists Ltd. Images (F--J) reproduced with minor adaption from Volkening and Sandstede \cite{volkening2} and licensed under CC-BY 4.0 (https://creativecommons.org/licenses/by/4.0/).}
\label{fig:bio}
\end{figure*}

Patterns are widespread in nature and often form due to the self-organization of independent agents. Whether exploring such collective dynamics in cancer \cite{ChoCancer}, wound healing \cite{WoundReview}, hair growth \cite{WangHair,GloverHair}, or skin pattern formation \cite{Jan,kondoTuringQuestion}, researchers focus on uncovering unknown cell behavior and signaling using a combination of experimental and modeling techniques. This process is complicated by the fact that biological patterns are inherently variable, making it challenging to quantify the distinguishing features of different mutants and judge model accuracy. In some applications, such as zebrafish skin patterns (Fig.~\ref{fig:bio}A--D), global information about patterns both \emph{in vivo} and \emph{in silico} is largely based on visual inspection, and this naturally leads to more subjectivity and limits the scale of the analyses. Moreover, the focus is often on the characteristic features of different mutants, making it unclear how much variability normally arises in mutant patterns, and how this variability compares to wild-type. To help address these challenges, here we develop a new methodology, based on topological data analysis and machine learning, for quantifying 
self-organized patterns with an automated, agent-based approach, and we apply our methods to study variability in zebrafish skin patterns.

Characterized by black and gold stripes, the zebrafish (\textit{Danio rerio}) is a model organism in the field of skin pattern formation \cite{McMen2016,Jan,irion2016chapter}. Remarkably, zebrafish stripes form due to the interactions of tens of thousands of different colored cells, which reliably self-organize on the growing skin despite their stochastic environment \cite{Yamaguchi,Maderspacher2003,ParTur130}. In addition to their namesake stripes, zebrafish feature a wealth of other patterns (e.g., spots and labyrinth curves \cite{Frohnhofer}) that form due to genetic mutations that restrict cell birth or alter cell behavior (often in unknown ways). While wild-type stripes (Fig.~\ref{fig:bio}A) are considered robust, mutants that lack certain cell types (Fig.~\ref{fig:bio}B--D) feature more variable spotty patterns \cite{Frohnhofer}. For example, the \emph{nacre} phenotype \cite{Maderspacher2003,Lister,Frohnhofer} has an enlarged central orange region with scattered blue splotches (Fig.~\ref{fig:bio}B). In comparison, both the \emph{pfeffer} \cite{Frohnhofer,Maderspacher2003,ParTur130,PatDev127} and \emph{shady} \cite{Lopes,Frohnhofer} mutants are characterized by dark spots, roughly aligned in stripes. These patterns differ in their finer details: \emph{pfeffer} has messy spots and peppered black cells across its skin, while \emph{shady} has sharp boundaries between light and dark regions \cite{Frohnhofer}. Although these descriptions apply in general, patterns vary due to the stochastic nature of pigment cell interactions.

Mathematical descriptions of zebrafish patterns
capture stochastic cellular interactions at different levels of detail. While partial differential equations (e.g., \cite{Yamaguchi,Nakamasu,Painter}) offer a broad perspective on the evolution of cell densities, cellular automaton \cite{Bullara,MorDeutsch} and agent-based models \cite{volkening,volkening2,Shinbrot} provide a more detailed view of individual cell behavior. For example, the agent-based model \cite{volkening2} specified cell interactions using stochastic rules to simulate zebrafish patterning \emph{in silico} (Fig.~\ref{fig:bio}F--I). Ideally, models should reproduce pattern formation as it is observed \emph{in vivo}, and this raises the question: how can we systematically quantify and compare pattern features, particularly in the presence of biologically-induced variability? Moreover, researchers seek to identify the cell interactions that are altered in mutant patterns, but this process is limited by the large number of parameters in agent-based models and the need for visual inspection to analyze simulation results. Reliable, automatic quantification of patterns (for both \emph{in vivo} and \emph{in silico} data) is therefore fundamental to measuring how well models perform and increasing their predictive potential. 

Many black-box machine learning algorithms have been developed for pattern classification, but interpretable methods provide results with a more transparent relationship to biological data. In this vein, Lee \emph{et al.} \cite{Lee2018} showed how to use ImageJ \cite{imageJ} to quantify some traits of giraffe spots; while their process can be automated, it relies on data in the form of contiguous blocks of bits in an image and only captures macroscopic pattern features, losing the underlying discrete, cell-based nature of the biological data. Taking a different approach, Miyazawa \emph{et al.} \cite{Miyazawa2010} assigned a ``pattern simplicity score'' (associated with the circularity of black-white boundary contours) to images of salmon patterns, and they quantified overall color tone by calculating the ratio of light to dark areas on fish images. These two global measures, which were applied to trout in \cite{Djurdjevic2019}, are broadly applicable but are not intended to capture detailed features. The methodology that we introduce in this paper, in contrast, utilizes the cell-based nature of skin patterns to quantify both macroscopic pattern attributes and microscopic features on the cellular level.

As shown in \cite{Topaz2015}, topological data analysis (TDA) has emerged as a valuable tool for characterizing collective behavior and self-organization. 
Tools from TDA, specifically persistent homology, allow one to assign shape descriptors to noisy or large data across a range of spatial scales. In the case of collective behavior, this translates to measuring topological summaries (e.g., connected components and loops) of the resulting patterns from the cellular level to the global level. In \cite{Topaz2015}, TDA was applied to study the velocity and positions of agents in simulations of a flocking model. By tracking global persistent homology features over time, Topaz \emph{et al.} \cite{Topaz2015} were able to identify agent clusters and detect the presence of global dynamics that would be challenging to notice visually. While such prior work \cite{Topaz2015,Miyazawa2010,Djurdjevic2019} has demonstrated how to quantify various overall features of patterns, 
characterizing the distinguishing traits of the different zebrafish patterns in Fig.~\ref{fig:bio} at the level of pigment cells requires a more detailed perspective.

Inspired by the utility of TDA for quantifying collective behavior, here we show how to reinterpret topological summaries as detailed measurements of pattern features. By combining TDA with interpretable machine learning techniques and working closely with the biological literature on zebrafish, we are able to automatically detect and quantify patterns given agent (e.g., cell) coordinate data. Our main contribution is an automated, interpretable framework for counting stripes and spots, detecting broken stripes, measuring stripe widths, quantifying stripe straightness, calculating spot size and roundness, measuring spot placement,
and estimating the onset of stripe formation from pattern data. To illustrate our techniques, we apply our methods to thousands of \emph{in silico} images of zebrafish patterns generated using the agent-based model from \cite{volkening2}. Because zebrafish display a wide range of patterns, we expect that our methodology can be applied to other problems in biological self-organization as well as to \emph{in vivo} data. 
Our approach opens up a range of possibilities for large-scale analysis of experimental images to better understand the cellular mechanisms underlying pattern formation.

\section*{Background and Methods}

Here we give a brief overview of zebrafish biology and the model \cite{volkening2}, as well as an introduction to the TDA and machine learning concepts that we use in our methodology (see the Supplementary Materials for additional background). 

\subsection*{Biological Background\label{sec:bio}}

Zebrafish stripe patterns consist of three main types of pigment cells: black melanophores, yellow/orange xanthophores, and silver/blue iridophores \cite{Frohnhofer} (Fig.~\ref{fig:bio}A). Xanthophores and iridophores are spread across the skin in two different forms (dense in light interstripes and loose in dark stripes), while black cells reside only in stripes \cite{ParTur130,hirata2005pigment,McMen2014,Mahalwar}. As these cells undergo differentiation, division, death, migration, and form changes, they self-organize into four to five stripes and four interstripes sequentially over a few months \cite{Jan}. Cells regulate each others' behavior through communication at short range (between neighboring cells) and at long range (between cells in stripes and interstripes) (e.g., \cite{Nakamasu,Patterson2013,Yamaguchi,2016heterotypic,PatNcomm}); see Fig.~\ref{fig:bio}E. Importantly, this regulation is inherently noisy. For example, cells may interact by reaching extensions toward their neighbors \cite{Inaba,delta,eom2015long}; whether or not cellular communication occurs then depends on if these extensions successfully find another cell.

Previously, models \cite{volkening,volkening2} have used estimates of wild-type stripe width \cite{delta,fadeev2016} and descriptions of developmental timelines (e.g., approximate times at which new stripes appear) \cite{Parichy,Jan,Singh} to judge model performance or fit parameters. Less data is available for zebrafish mutants,
and, to our knowledge, global information is in the form of qualitative descriptions of the characteristic features of their patterns. Local measurements, in turn, include cell speeds \cite{TakahashiMelDisperse,Yamanaka2014} and distances between adjacent cells \cite{TakahashiMelDisperse,2016heterotypic,ParTur256}. Notably, we are not aware of measurements of pattern variability or stripe straightness.

\subsection*{Model and Generation of \emph{In Silico} Pattern Data\label{sec:model}}

The model \cite{volkening2} treats pigment cells as individual agents (point masses) and tracks their positions (namely $(x,y)$-coordinates) in space as they interact on growing $2$--D domains. The behavior of five different types of cell agents is accounted for: we let $\textbf{M}_i(t)$ be the $(x,y)$-coordinate of the $i$th melanophore ($M$) at time $t$; similarly, $\textbf{X}^d_i(t)$, $\textbf{X}^\ell_i(t)$, $\textbf{I}^d_i(t)$, and $\textbf{I}^\ell_i(t)$ denote the locations of the $i$th dense xanthophore ($X^\text{d}$), loose xanthophore ($X^\text{l}$), dense iridophore ($I^\text{d}$), and loose iridophore ($I^\text{l}$), respectively; see Fig.~\ref{fig:bio}K. Space is continuous, and cell movement, which includes repulsion and attraction, is modeled by coupled ordinary differential equations. Cell birth, death, and transitions in type, in turn, take the form of stochastic, discrete-time rules. These rules, which are strongly motivated by the biological literature (e.g., \cite{Nakamasu, Frohnhofer,PatNcomm,TakahashiMelDisperse}), depend on the number of cells in disk and annulus neighborhoods centered at the cell or location of interest (see Fig.~\ref{fig:bio}J). The authors \cite{volkening2} use these neighborhoods to model the cells that a given cell (or precursor) could communicate with (e.g., through direct contact \cite{walderich2016homotypic}, diffusing substances \cite{PatNcomm}, or 
dendrite extensions \cite{Inaba,delta,eom2015long} as in Fig.~\ref{fig:bio}E). As an example cell interaction rule, interstripe cells are known to promote $M$ differentiation at long-range \cite{Patterson2013,Nakamasu}, and these dynamics are modeled as follows:
\begin{align}
&\frac{\sum_{i=1}^{N^\text{d}_\text{X}}{ \mathds{1}}_{\Omega^{\textbf{z}}_{\text{long}}} (\textbf{X}^\text{d}_i) + \sum_{i=1}^{N^\text{d}_I}{ \mathds{1}}_{\Omega^{\textbf{z}}_{\text{long}}} (\textbf{I}^\text{d}_i)}{\alpha + \beta \sum_{i=1}^{N_\text{M}}{ \mathds{1}}_{\Omega^{\textbf{z}}_{\text{long}}}(\textbf{M}_i)} > 1 \nonumber   \\
& \Longrightarrow \hskip0.3cm \text{$M$ birth at $\textbf{z}$ (if not overcrowded)}, \label{rule:Mbirth}
\end{align}
where $\textbf{z}$ is a randomly selected location to be evaluated for possible cell birth; $N^\text{d}_\text{X}$, $N^\text{d}_\text{I}$, and $N_\text{M}$ are the numbers of $X^\text{d}$, $I^\text{d}$, and $M$ cells on the domain, respectively; and $\Omega_\text{long}^\textbf{z}$ is an annulus centered at $\textbf{z}$ that models long-range cellular communication (see Fig.~\ref{fig:bio}J). According to \eqref{rule:Mbirth}, a new $M$ cell appears at position $\textbf{z}$ when the ratio of interstripes cells to $M$ cells at long-range is greater than one. (Note that the interaction rules in \cite{volkening2} are given in terms of numbers, rather than proportions, of cells. We have adjusted the model \cite{volkening2} so that these rules depend on the ratios or densities of cells in different regions, as this framework works better for our large-scale study; see the Supplementary Materials  for more details.)

The agent-based model \cite{volkening2} can be used to simulate the full timeline of adult pattern formation from when it begins when the fish is roughly $21$ days post fertilization (dpf). Because the model \cite{volkening2} is stochastic, simulating it repeatedly leads to different \emph{in silico} patterns and, importantly, for our methods, cell-coordinate data. We thus generate an extensive data set by simulating the development of thousands of zebrafish patterns. 
We simulate wild-type development from $21$~dpf until $66$ dpf, at which point zebrafish are expected to have three complete interstripes, two complete stripes, and some partially-formed stripes near the boundaries. We simulate \textit{nacre} and \textit{pfeffer} pattern formation until $76$ dpf and \textit{shady} development until $96$ dpf by turning cell birth off for the appropriate cell types as described in \cite{volkening2}. (We note that experimentalists often use 
stages \cite{Parichy} rather than dpf to measure time; in the 
model \cite{volkening2}, $66$ dpf, $76$ dpf, and $39$--$44$ dpf correspond to the Juvenile, Juvenile+, and Squamation onset Posterior stages, respectively.) With one exception, we perform all of our analyses on the final simulated patterns at $66$ dpf (for wild-type), $76$ dpf (for \textit{nacre} and \pfeffer), and  $96$ dpf (for \shady). Following the approach in \cite{volkening2}, we enforce periodic boundary conditions in the horizontal direction and wall-like boundary conditions at the top and bottom of the domain (see Fig.~\ref{fig:method_example}A). To help avoid quantifying partially-formed stripes or spots, we remove the cells in the top and bottom 10\% of the domain in post-processing. 

To generate our first data set, we simulate wild-type, \emph{nacre}, \emph{pfeffer}, and \emph{shady} patterns under the baseline conditions and parameters described in \cite{volkening2}. We then adjust the model to account for more realistic biological stochasticity in cell interactions. In particular, rather than using deterministic length scales in the cell interaction rules, each day we select these length scales 
randomly per cell and interaction from a normal distribution centered at the default parameter value. In our last data set, we focus on the inner radius of $\Omega_\text{long}$ in \eqref{rule:Mbirth} and explore the role of
this parameter 
while keeping all other parameters at their default values.

\subsection*{Topological Data Analysis and Machine Learning\label{sec:TDA}} 

\begin{figure}[t!]
    \centering
    \includegraphics[width=0.47\textwidth]{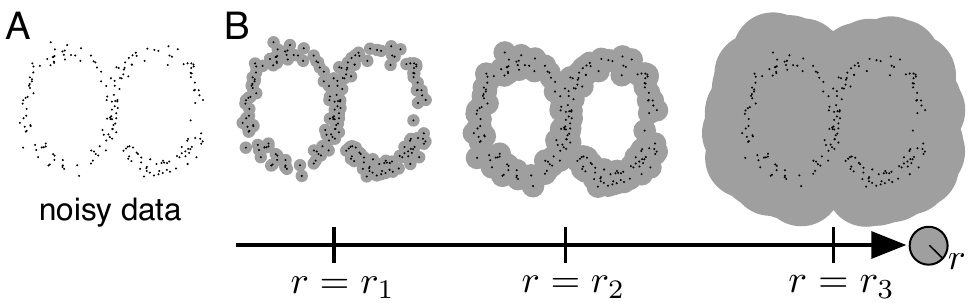}
    \caption{Illustration of persistent homology applied to coordinate data. (A) Noisy data sampled from a figure-eight shape and (B) corresponding manifold expansions.}
     \label{fig:Figure8_example}
\end{figure}

Our approach to quantifying 
patterns relies on topological data analysis and machine learning. Topological data analysis (TDA) is an emerging branch of mathematics and statistics that aims to extract quantifiable shape invariants from complex and often large data \cite{Carlsson2009,chazal, Edelsbrunner10, Ghrist2014, Zomorodian2009}. One of the main tools in TDA is known as persistent homology, which we review now briefly. Given a data set of $N$ discrete points $\{\textbf{x}_i\}_{i =1}^N$ that lie in some metric space $(D, d_D)$, we place a ball of radius $r$ 
at each $\textbf{x}_i$ to obtain the set $b_r(\textbf{x}_i) = \{\textbf{y} \in D : d_D(\textbf{x}_i, \textbf{y}) \leq r\}$.
We then take the union of these balls over all $i \in [1,N]$,
namely $\bigcup_{i \in [1,N]} b_r(\textbf{x}_i)$. This process yields a new manifold with shape generated by the original data, and persistent homology tracks how the shape of this manifold changes as $r$ increases (see Fig.~\ref{fig:Figure8_example}).

 For our work it suffices to view the dimension $0$ and dimension $1$ persistent homology groups as vector spaces whose dimensions correspond the number of connected components and loops, respectively, of the evolving manifold (see the Supplementary Materials and \cite{Carlsson2009,chazal, Edelsbrunner10, Ghrist2014, Zomorodian2009} for more details and definitions). The number of generators of the $i$th homology group is called the $i$th betti number, denoted $\beta_i$. If a topological feature (e.g., connected component or loop) appears at some radius $r_b$ and disappears at some radius $r_d > r_b$, then we say this feature is born at  $r = r_b$ and dies at $r = r_d$, and its persistence is given by $r_d-r_b$. 

For example, because a figure eight has one connected component and two loops, this shape has $\beta_0  = 1$ and $\beta_1  = 2$. Now consider a noisy data set sampled from a figure eight, as we show in Fig.~\ref{fig:Figure8_example}A. To compute the persistent homology of this data we take the union of balls of radius $r$ centered around each data point for an increasing sequence of $r$ values. Two loops appear in the data at $r = r_2$ and disappear before $r = r_3$ in Fig.~\ref{fig:Figure8_example}B, so this data set has two dimension $1$ homology generators that are both born at $r_b = r_2$ and die at $r_d = r_3$ (with persistence given by $r_3-r_2$).
Similarly, this data set is connected for $r \geq r_2$, so it has one dimension $0$ homology generator for $r \geq r_2$ with infinite persistence and several dimension $0$ homology generators for $r < r_2$. Thus, persistent homology reveals that the noisy data in Fig.~\ref{fig:Figure8_example}A is topologically similar to a figure-eight shape ($\beta_0  = 1$, $\beta_1  = 2$) for $r_2 \leq r < r_3$.

In addition to using TDA, we apply methods from interpretable machine learning to quantify patterns. Machine learning algorithms seek to automatically learn information from a given data set for classification or prediction purposes \cite{Bishop2006, HastieBook}. The machine learning approach we use involves clustering data into different classes based on a similarity measure. Specifically, we apply single-linkage clustering to subsets of agents (e.g., pigment cells) to identify clusters corresponding to spot or stripe patterns. Single-linkage clustering is an agglomerative hierarchical clustering method: each data point begins in its own cluster and points (or clusters of points) are merged sequentially based on which two clusters are closest to each other \cite{Bishop2006, HastieBook}. We continue this process until there are $n$ clusters, where $n$ is either one or some predetermined number of desirable clusters. We use single-linkage clustering over other clustering algorithms (e.g., average linkage or k-means) to capture elongated, undulating, and non-spherical clusters that are characteristic of some zebrafish mutants (see Fig.~\ref{fig:bio}).

As a side note, dimension $0$ persistent homology is analogous to single-linkage clustering, so there is a natural connection between TDA- and clustering-based methods for pattern quantification  \cite{Carlsson2009}. Using clustering and topological methods in tandem yields both multi-dimensional, coordinate-free summaries (from TDA) and essential information about the locations of different agents (from clustering).

\section*{Results: Our Methodology for Quantifying Patterns}

We now use TDA and machine learning to develop 
our main result: an interpretable, agent-based
methodology for automatically quantifying patterns that arise due to self-organization. We summarize our methods in Table S1 and illustrate how they can be applied to zebrafish in Fig.~\ref{fig:method_example}. Direct methods for measuring local pattern features are further presented in the Supplementary Materials.

Tailored to a specific application (zebrafish), our work opens up a new way of thinking about TDA tools and utilizing them to obtain detailed measurements of patterns. We expect that a similar approach can be used to study other patterns with data in the form of agent coordinates or images (with functional persistence). To help encourage further applications of TDA to self-organized patterns, we thus present our methods using general language in the next section, while also using zebrafish to highlight the kinds of application-specific considerations one must address when applying TDA to new data.
In particular, one application-specific step involves determining what agent type(s) to use as input for topological feature computations. For example, multiple types of cells are present in the same pattern features on zebrafish (e.g., in Fig.~\ref{fig:bio}A, both $I^\text{d}$ and $X^\text{d}$ appear in interstripes). Applying TDA to the locations of every agent type in a pattern is expensive. It may be sufficient to study only one or two agent types, but selecting which types to use requires application-specific considerations.

\begin{figure*}[h!]
    \centering
    \includegraphics[width=\textwidth]{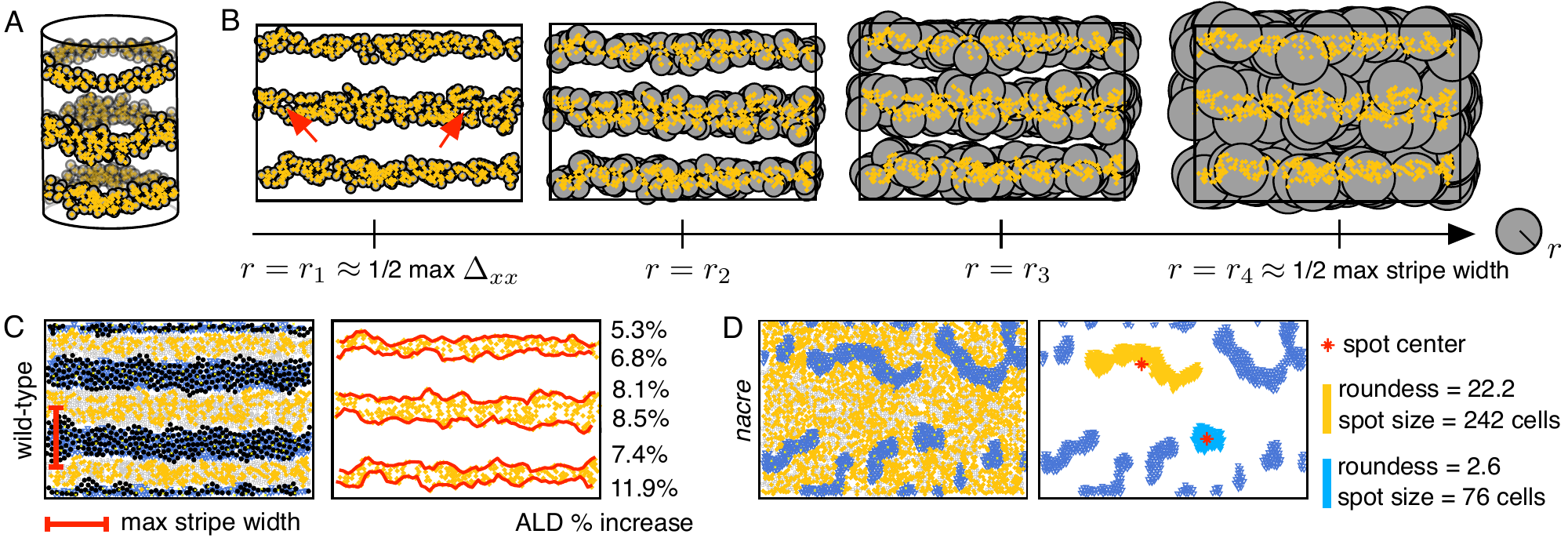}
    \caption{Illustration of our topological techniques applied to zebrafish patterns. (A) Boundary conditions are periodic in the $x$-direction, so stripes and interstripes are viewed as loops from a topological perspective. (B) We count interstripes and measure stripe width using persistent homology. We show manifold expansions of the locations of $X^\text{d}$ cells by considering balls of growing radius $r$ centered at the location $\textbf{X}^\text{d}_i$ of each cell. When $r = r_1$, the radius of the balls is about half the maximum distance between neighboring $X^\text{d}$ cells $\Delta_{xx}$. At this point, three interstripes have formed, but the number of loops is larger than the true number of interstripes due to gaps between cells, highlighted by red arrows ($\beta_0 = 3$ and $\beta_1 > 3$). As $r$ increases to $r_2$, the noisy loops die off, leaving only three loops ($\beta_0 = 3$ and $\beta_1 = 3$). The long persistence  of three loops corresponds to the true presence of three interstripes. As $r$ increases further to $r_4$, the manifold collapses to a single connected component ($\beta_0 = 1$ and $\beta_1 = 1$). The difference between the ball radius at which this collapse occurs ($r_4$) and the ball radius at which three loops appear ($r_1$) approximates half the maximum width of black stripes. (C) By combining TDA with clustering methods, we automatically detect interstripe boundaries and measure their curviness; we show the percent increase in arc length distance (ALD) of these boundaries (traced out in red) relative to perfectly straight stripes here. (D) We describe spotted phenotypes by combining persistent homology, clustering methods, and principal component analysis. We use $\beta_0$ to quantify the number of spots. As an example, we show the spot size and spot roundness for two \emph{nacre} spots. }
     \label{fig:method_example}
\end{figure*}

\subsection*{Counting Spots and Stripes\label{sec:toolscount}}

We compute the dimension $0$ and dimension $1$ persistent homology groups using the coordinate data of agents (e.g., pigment cell locations generated by the model \cite{volkening2}) to quantify pattern types, assuming periodic boundary conditions in the $x$-direction. With these boundary conditions, spots can be viewed as connected components without loops, whereas stripes wrap around the domain and are thus connected components with a single loop (see Fig.~\ref{fig:method_example}A--B). Consequently, $\beta_0$ and $\beta_1$ approximate the number of spots and stripes in a pattern, respectively.\footnote{If boundary conditions are not periodic, spots and stripes are topologically equivalent.}

For zebrafish, we estimate the number of stripes and interstripes in wild-type patterns by computing $\beta_1$ for $X^\text{l}$ and $X^\text{d}$ cells, respectively. We apply TDA to these cells because they uniformly cover the fish skin, but in different forms in stripes and interstripes.\footnote{Alternatively, we could have used $I^\text{d}$ and $I^\text{l}$, but these cells appear closer together than xanthophores at stripe--interstripe boundaries.} 
 We estimate the number of spots in \emph{nacre} and \emph{pfeffer} patterns by computing $\beta_0$ using the locations of blue $I^\text{l}$ cells.\footnote{Alternatively, we could have computed the number of spots using $X^\text{l}$ in \emph{nacre}, as these cells appear together with $I^\text{l}$. However, $X^\text{l}$ are much less dense than $I^\text{l}$ in \emph{nacre}, so the former would introduce more noise into our measurements.
} For \emph{pfeffer}, individual $M$ cells appear randomly
on the domain, so using these cells to count the number of spots would introduce spurious connected components (in the form of individual black cells). In comparison, $M$ are much more clustered in \emph{shady}; thus, we calculate the number of dark \emph{shady} spots by computing $\beta_0$ for $M$. 

In general, we calculate betti numbers by applying persistent homology to the agents' coordinate data and using a persistence threshold to count the number of homological generators whose persistence is greater than the set threshold ($T_p$). Empirical estimates of cell--cell spacing motivate our choice of $T_p$ for zebrafish. Specifically, we use $T_p^0= 100$~$\mu$m and $T_p^0 = 90$~$\mu$m as the dimension $0$ persistence thresholds for iridophores and melanophores, respectively. We chose these thresholds conservatively, as average xanthophore--xanthophore neighboring distances are $30$--$60$~$\mu$m and average melanophore--melanophore distances are roughly $50$--$60$ $\mu$m in wild-type \cite{TakahashiMelDisperse,ParTur256,2016heterotypic,volkening2}. (We are not aware of empirical measurements of iridophore spacing.)
For dimension $1$ homology, we use a universal persistence threshold of $T_p^1 = 200$~$\mu$m. Moreover, to ensure that we correctly differentiate between complete and broken stripes or interstripes, we specify that a persistence generator only counts toward $\beta_1$ if its birth radius $r_b$ is below a certain threshold ($T_b^1$).  
For $X^\text{l}$ and $X^\text{d}$, we use $T_b^1 = 100$ $\mu$m and $T_b^1 = 80$ $\mu$m, respectively. These thresholds were motivated by empirical cell--cell distance measures \cite{TakahashiMelDisperse,ParTur256,2016heterotypic} and tuned based on parameter fitting experiments with stripe and interstripe breaks.

Simultaneously, we can use persistent homology to identify stripe breaks when the number of expected stripes is known (see Fig.~\ref{fig:default_WT}A for examples of stripe and interstripe breaks). 
Namely, we flag a stripe break 
when $\beta_1$ is less than the expected number of stripes. Here we additionally consider $\beta_1$ of the $M$ cells, with $T_b^1 = 90$ $\mu$m and $T_p^1 = 200$ $\mu$m. We compute $\beta_1$ for both $X^\text{l}$ and $M$ because the former appear at low density in dark stripes; computing $\beta_1$ for both cell types allows us to be more confident in our results. 
As we discussed in \textit{Background and Methods}, we expect that our simulated zebrafish patterns
have two fully formed stripes and three fully formed interstripes at the time of our analysis, so we flag a stripe break when $\beta_1 < 2$ for both 
$X^\text{l}$ and 
$M$ cells. Similarly, we flag an interstripe break when $\beta_1 < 3$ for 
$X^\text{d}$ cells.

\subsection*{Measuring Stripe Width\label{sec:tools-SW}}

Beyond quantifying the number of stripes or spots, we leverage TDA to approximate stripe and interstripe widths. In particular, we estimate (inter)stripe widths using the persistence ($r_d-r_b$) of the significant dimension $1$ persistence points. We define significant dimension $1$ persistence points
as those with persistence greater than or equal to $T_p^1$ and birth radius $r_b$ less than or equal to $T_b^1$. For example, the persistence of a stripe loop 
is the difference between the radius value ($r_d$) at which two adjacent 
stripes combine to form a single loop and the radius value ($r_b$) at which the 
stripe feature initially formed (here we ignore the features that persist to infinity). This difference ($r_d-r_b$) corresponds to half of the maximum distance between two adjacent stripes, capturing the maximum width of the enclosed interstripe 
(see Fig.~\ref{fig:method_example}). 

In wild-type zebrafish, twice the persistence of the yellow $X^\text{l}$ loops yields an approximation for the maximum interstripe width across the fish. Similarly, twice the persistence of the orange $X^\text{d}$ loops approximates an upper bound on stripe width. We note that $r_d$ alone could be used as an alternative estimate for maximum (inter)stripe width, but we use $r_d - r_b$ to account for the narrow boundary region between stripes and interstripes on zebrafish. To obtain a lower bound on stripe width, one could calculate the persistence of the significant dimension $0$ persistence points, as this measurement is based on half of the minimum distance between two adjacent interstripes.

\subsection*{Measuring Spot Size\label{sec:tools-SpS}}

We measure spot size by applying single-linkage hierarchical clustering to the agents of interest with the number of desired clusters (e.g., number of spots) set to the $\beta_0$ values we obtained from our topological analyses. Then, we count the number of cells per cluster to approximate the size of each spot. 
We define ``spot size'' as the median number of agents per spot across all of the spots.

\subsection*{Quantifying Stripe Straightness\label{sec:tools-SS}}

To measure ``stripe curviness'' we compute the arc length distance (ALD) of the boundary of each single-linkage cluster that corresponds to a stripe. We define our stripe curviness measure to be the average percent increase of this ALD from the ALD of straight stripes:

\begin{align} \text{curviness} =  \underset{\text{stripes}}{\text{mean}}\left( \left( \frac{\text{true ALD}}{\text{straight ALD}} - 1\right) \times 100  \right). \label{eq:curviness}
\end{align}
For example, to measure the curviness of wild-type zebrafish stripes, we apply single-linkage clustering to the locations of $X^\text{d}$ cells. For the number of desirable clusters $n$, we use the number of expected interstripes minus the number of stripe breaks that we identified with persistent homology (see Fig.~\ref{fig:method_example}C). We then calculate the ALD for the resulting clusters and compute stripe curviness using \eqref{eq:curviness}.

\subsection*{Quantifying Spot Roundness\label{sec:tools-SR}}
To estimate spot uniformity, we use the clusters identified via single-linkage hierarchical clustering (with the number of desired clusters set to the $\beta_0$ values). We then apply principal component analysis (PCA) to each cluster. The eigenvalue decomposition in PCA provides information about how varied the data is in each dimension. Since our data is two-dimensional, we use PCA to evaluate the spread of each cluster in the $x$- and $y$-directions. If a spot has significantly more variance in one direction, this indicates that it is irregularly shaped or elongated. Specifically, we define our roundness measure as:
\begin{align} \text{roundness of spots} &= \underset{\text{spots}}{\text{median}} \left( \frac{\text{PCA eigenvalue 1}}{\text{PCA eigenvalue 2}} \right ). \label{eq:PCA}
\end{align}
We assume that a PCA eigenvalue ratio close to one implies round spots, while a PCA eigenvalue ratio $\gg 1$ indicates irregular, non-uniform spots (see Fig.~\ref{fig:method_example}D for examples).

\subsection*{Determining Spot Alignment and Center Width\label{sec:tools-SACR}}

We quantify spot alignment by first applying single-linkage hierarchical clustering to agent locations (with the number of desired clusters set to the $\beta_0$ values). We then calculate the pairwise $l_{\infty}$ distances between the cluster centroids and complete a nearest-neighbor search with the $l_{\infty}$ metric.\footnote{Note that $d_{l_{\infty}}((x_1, y_1), (x_2, y_2)) = \max(|x_1 - x_2|, |y_1 - y_2|)$.} 
This allows us to extract the distance from each spot to its closest neighboring spot. We define the spot-spacing variance as the standard deviation of these nearest-neighbor $l_{\infty}$ distances. A large spacing variance corresponds to non-uniform spot placement, while a low spacing variance predicts well-aligned spots. 

Motivated by the \emph{nacre} and \emph{shady} patterns, which feature expanded light central regions (see Fig.~\ref{fig:bio}G and Fig.~\ref{fig:bio}I), we also use the cluster centroids to approximate the center width, defined as twice the distance from the midpoint of the domain to its first spot.
In particular, we estimate the center width as twice the minimum distance from the cluster centroids to the midpoint of the domain, minus the median spot diameter. Here we define spot diameter 
as twice the greatest Euclidean distance from the spot's centroid to cells belonging to the spot. For zebrafish, the center radius corresponds to the width of the central interstripe X0; see Fig.~\ref{fig:bio}F--G.

\subsection*{Capturing Pattern Formation Events\label{sec:tools-PFE}}

Thus far, we have focused on quantifying pattern features at a snapshot in time. However, for self-organization that occurs during organism development, it is also useful to estimate the time at which specific features emerge. For example, in wild-type zebrafish, the second and third interstripes X1V and X1D (see Fig.~\ref{fig:bio}F) 
develop around $39$--$44$ dpf (based on approximations \cite{volkening2} of images in \cite{Frohnhofer,Parichy}). This information on target time dynamics serves as an additional quantitative measurement that can be used to evaluate models. Here we present a method for quantifying the time 
at which stripes X1V and X1D form; future work could extend these methods to capture the time dynamics of spot formation and other features.

Given data in the form of agent locations at consecutive time points, we first assume new stripes form somewhere between day $d_0$ and $d_1$. If there is no prior knowledge about the expected time of stripe development, one can set $d_0$ and $d_1$ to the first and last days of pattern development, respectively. For zebrafish, because the model \cite{volkening2} was parameterized so that interstripes X1D and X1V form around $39$--$44$ dpf, we conservatively set $d_0 = 32$ dpf and $d_1 = 62$ dpf. Within the specified time interval, we then analyze the patterns sequentially beginning at $d_0$, assuming there is initially a single stripe on the domain. At each time step, we find the upper and lower bounds of the stripes by computing the maximum and minimum, respectively, of the $y$-coordinates of the agents of interest (e.g., for zebrafish, we use $X^\text{d}$ cells). 
Finally, we estimate the initial formation of new stripes as the first day at which the upper or lower bounds of the stripes increase by more than some threshold from the previous day. For zebrafish, the threshold we use is 200 $\mu$m.\footnote{Alternatively, we could rely on topological summaries to approximate the initial formation of new stripes, but a direct approach is more computationally efficient in this setting.}

\section*{Results: A Quantitative Study of Zebrafish Patterns \label{sec:ZFresults}} 

We now study zebrafish pattern variability and robustness by analyzing thousands of \emph{in silico} wild-type and mutant patterns generated using the agent-based model \cite{volkening2}. Quantitatively evaluating data of this scale is possible because of our automated framework. As a baseline test, we begin by illustrating our techniques on simulations of wild-type zebrafish stripes. Because our analysis there is consistent with previous characterizations collected visually and local pattern measurements, we then use our methods to extract quantifiable features from mutant patterns and measure pattern variability in the presence of increased stochasticity in cell interactions. We conclude by showing how our methods can be used to 
detect the impact of changing a given model parameter without the 
need for visual inspection.

We view our results in the next sections as presenting a broader, more objective picture of the behavior of the agent-based model \cite{volkening2}. Additionally, because this model is closely based in the biological literature, our results serve to predict the kind of pattern variability we expect to see \emph{in vivo} based on the model \cite{volkening2}. As large-scale collections of experimental images become available, our predictions can be tested by applying our techniques to \emph{in vivo} 
images of zebrafish as well.

\subsection*{Illustrating our Techniques on Wild-type Zebrafish\label{sec:default}}

\begin{figure}[t!]
    \centering
    \includegraphics[width=0.44\textwidth]{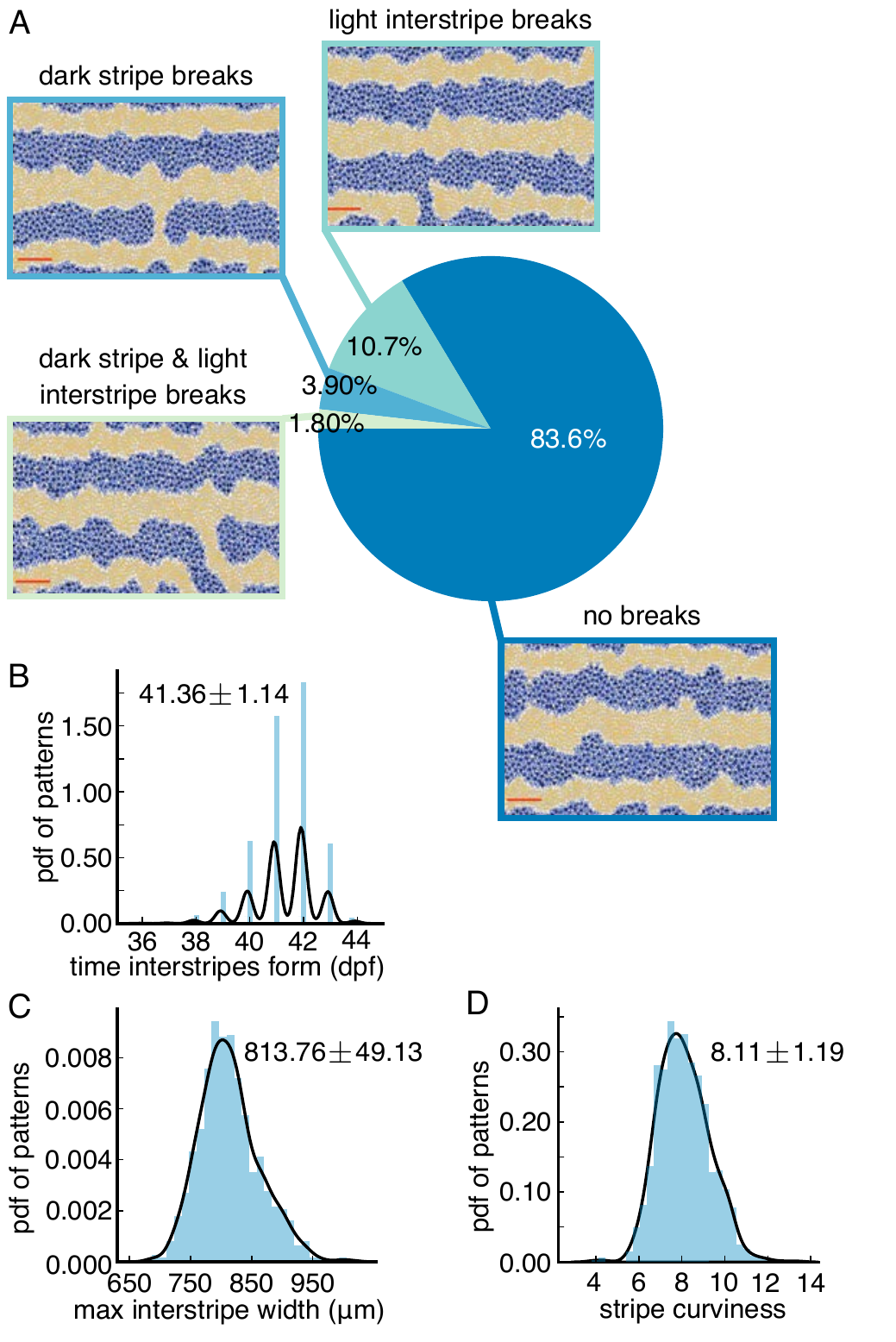}
    \caption{Baseline quantification of wild-type patterns. All measurements are based on $1,000$ simulations of the model \cite{volkening2} under the default parameter regime. (A) We use persistent homology to detect the presence of breaks in stripes and interstripes. (Following the example in \cite{volkening2}, we do not count breaks in the dark stripes along the top and bottom boundaries of the domain.) Red scale bar is $500$ $\mu$m. (B) Distribution of times at which interstripes X1D and X1V (see Fig.~\ref{fig:bio}F) begin to form. (C) Distribution of maximum interstripe width. (D) Distribution of stripe curviness; also see Fig.~\ref{fig:method_example}C. In (B--D), we display histograms of \emph{in silico} data and kernel density estimator (KDE) curves with a Gaussian kernel in black; the mean plus/minus the standard deviation is
    shown in each plot for the data.} 
    \label{fig:default_WT}
\end{figure}

We focus on stripes first because they provide a means of testing our methodology, as wild-type patterns have the most experimental data (collected both \emph{in silico} and \emph{in vivo}) available for comparison. Here we use our methodology to evaluate $1,000$ wild-type zebrafish patterns generated with the model \cite{volkening2} under the default parameter regime. Previously, model performance \cite{volkening2} was judged by manually counting the number of stochastic simulations that display breaks (or interruptions) in interstripes and requiring matches in pattern features (e.g., number of interstripes present) at major developmental timepoints.
In particular, by inspecting $100$ \emph{in silico} patterns, the authors \cite{volkening2} reported a success rate of $89$\% according to the former goal, meaning that $89$ out of $100$ simulations had no interruptions in interstripes. (Note that breaks in black stripes are occasionally seen on real fish, so these interruptions were not quantified in \cite{volkening2}.) Our methodology allows us to analyze much larger data sets and remove any human error from the process; we demonstrate how topological methods can be used to detect stripe breaks automatically in Fig.~\ref{fig:default_WT}A. Across $1,000$ wild-type simulations, we find that $87.5$\% have no breaks in interstripes (flagged by a decrease in $\beta_1$ for  $X^\text{d}$ cells). This agrees well with the success rate in \cite{volkening2} that was computed using visual inspection.

As an additional evaluation measure, we manually viewed $200$ model outputs and found that the betti numbers capture interstripe breaks with $100$\% accuracy and only one false positive. In a similar vein, the model \cite{volkening2} was parameterized so that interstripes X1D and X1V (see Fig.~\ref{fig:bio}F) form between $39$--$44$ dpf, but until now this property was judged by visual inspection. Using our automated methods, we show the distribution of times at which these interstripes develop in Fig.~\ref{fig:default_WT}B and find good agreement with the target pattern milestones 
in \cite{volkening2}.

Fig.~\ref{fig:default_WT}C--D show the distributions of interstripe width and stripe curviness across $1,000$ wild-type simulations. The maximum interstripe width, measured by the persistence of the significant dimension $1$ persistence points of $X^\text{l}$, represents the maximum separation between adjacent stripes. We find that this quantity has a mean of about $814$~$\mu$m and a standard deviation of approximately $49$ $\mu$m, which is similar to the average distance between cells \cite{TakahashiMelDisperse,ParTur256}, suggesting that variance in stripe width may be caused by the addition or loss of one cell.  Similarly, in Fig.~\ref{fig:default_WT}D, we show measurements of wild-type stripe curviness \eqref{eq:curviness}, a dimensionless quantity that could be compared to empirical data in the future. More generally, Fig.~\ref{fig:default_WT}B--D provide a baseline measurement of the model output \cite{volkening2} that we use to compare to further studies.

\subsection*{Quantifying ``Characteristic'' in Noisy Mutant Patterns\label{sec:mutants}}

\begin{figure}[t!]
    \centering
    \includegraphics[width=0.44\textwidth]{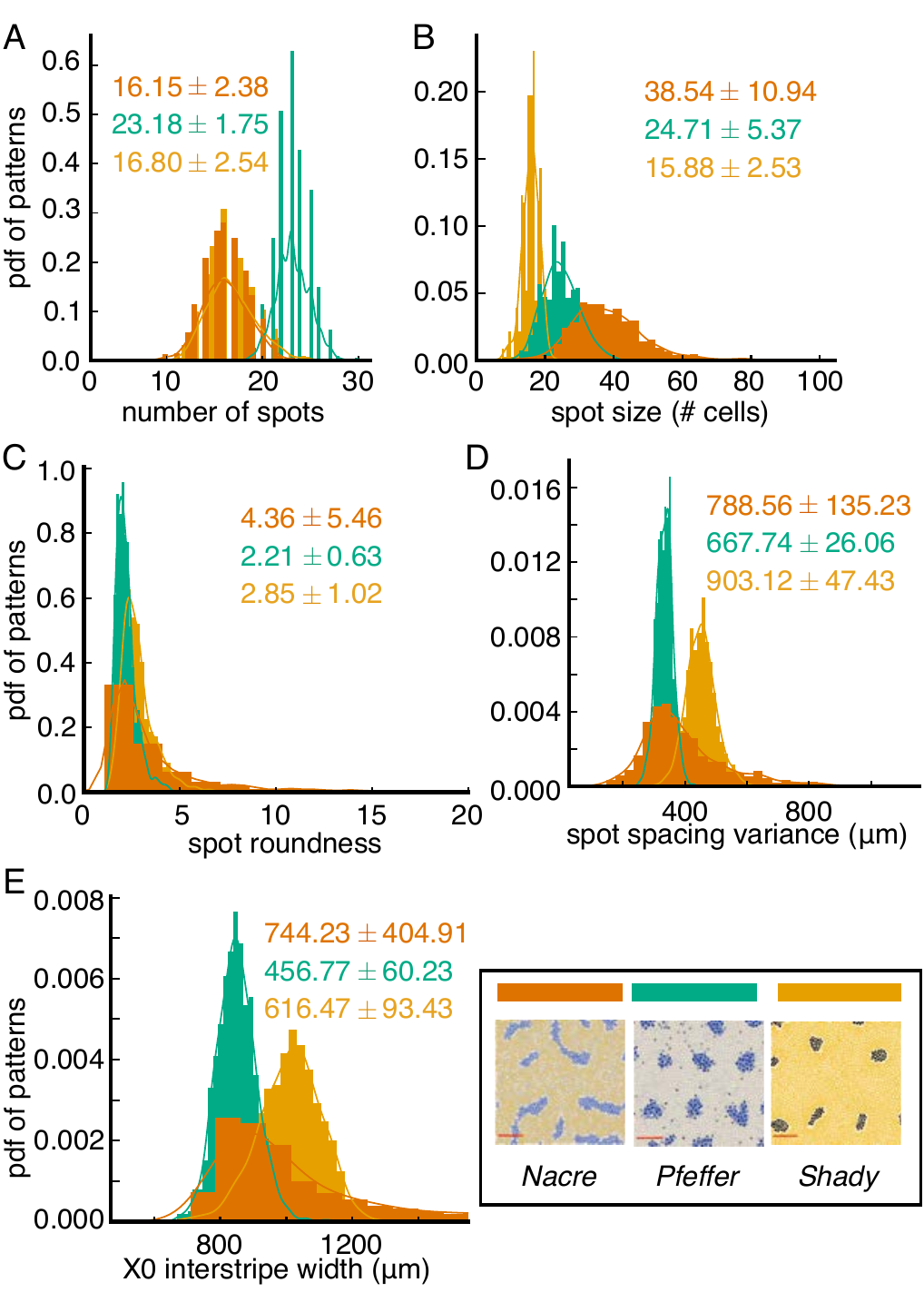}
    \caption{Baseline study of mutant patterns to extract quantifiable features. All measurements are based on $1,000$ simulations of the model \cite{volkening2} (for each mutant) under the default parameter regime. Histograms show distributions for (A) the number of spots, (B) spot size, (C) spot roundness, (D) variance in spot spacing, and (E) X0 interstripe width (see Fig.~\ref{fig:bio}G). We show KDE curves with a Gaussian kernel overlaid on the histograms of data; the mean plus/minus the standard deviation is  shown in each plot for the data.
} 
    \label{fig:default_ESM_hists}
\end{figure}

The \emph{nacre}, \emph{pfeffer}, and \emph{shady} mutants lack specific cell types, 
leading to altered patterns, which are highly variable and can be broadly described as spotty (see Fig.~\ref{fig:bio}B--D). Here we use our methods to analyze $1,000$ \emph{in silico} patterns generated with the model \cite{volkening2} under the default parameter regime for each mutant. Our results, shown in Fig.~\ref{fig:default_ESM_hists}, serve as quantitative descriptors of what constitutes ``characteristic'' for each mutant (according to the model) and demonstrate our methods' abilities to extract quantifiable differences between spot patterns. Among the three mutants, we find that \emph{pfeffer} has the most spots and that these spots are the most round and the most evenly spaced (Fig.~\ref{fig:default_ESM_hists}A, C--D). In comparison, \emph{nacre} and \emph{shady} have a similar number of spots, but the spots on \emph{shady} are smaller and rounder than those of \emph{nacre}.
(As we noted in \emph{Background and Methods}, we remove a small region at the top and bottom of the domain prior to our analysis to avoid quantifying partial spots.) Moreover, the width of the central X0 interstripe in \emph{pfeffer} is closest to wild-type interstripe width (see Fig.~\ref{fig:default_WT}C), while both \emph{nacre} and \emph{shady} feature expanded central interstripes, echoing empirical observations \cite{Frohnhofer}. Interestingly, we find that the variance in the number of spots for all three mutants is small (a standard deviation of about two spots). With the exception of \emph{nacre}, which displays the greatest variability in four of the five measurements we present in Fig.~\ref{fig:default_ESM_hists}, the variance in spot spacing and the width of the central interstripe X0 is also small (on the order of the distance between neighboring cells \cite{TakahashiMelDisperse}). In the future, it would be interesting to compare these quantities to large-scale \emph{in vivo} data and determine what cell interactions in the model \cite{volkening2} are responsible for selecting them robustly.

\subsection*{Measuring Pattern Variability\label{sec:LSnoise}}

Some cellular interactions on the zebrafish skin are thought to be regulated by direct contact, dendrites, or longer projections \cite{Inaba,delta,eom2015long} (see Fig.~\ref{fig:bio}E). To account for this,
the model \cite{volkening2} assigns disk (short-range communication) and annulus (long-range communication) interaction neighborhoods to each cellular agent (see Fig.~\ref{fig:bio}J). 
Cell birth, death, and form transitions are then governed by rules (e.g., \eqref{rule:Mbirth}) that depend on the proportion of cells within these neighborhoods. The size of the neighborhoods dictates which cells are able to interact and therefore plays a critical role in patterning. While the interaction neighborhoods have deterministic sizes (based on empirical measurements \cite{TakahashiMelDisperse, delta, eom2015long}) in \cite{volkening2}, a more realistic model should account for stochastic variations in cell size and projection length. Randomly varying the length scales involved in the interaction neighborhoods serves as a means of including more realistic cellular communication (which could also include diffusion of signaling factors \cite{PatNcomm} in the future) in agent-based models of zebrafish. As a first step toward including more realistic stochasticity, we therefore replace the deterministic length scales
in the model \cite{volkening2} with stochastic length-scale parameters and measure their effect on pattern variability. This models the presence of randomness in cell interactions due to variations in cell size and projection length.

Interaction neighborhoods appear in $17$ places in the rules that govern $M$ birth, $M$ death, iridophore form changes, and xanthophore form changes in the model \cite{volkening2}. For each cell interaction, we randomly select the size of the associated interaction neighborhood per cell per day from a normal distribution with the mean set to the default parameter value.
We vary the standard deviation from $1$--$50$\% of the mean and for each standard deviation (we consider $\sigma \in \{0.01,0.05,0.1,0.2,0.3,0.5\}$, where $\sigma$ times the default length scale is the standard deviation of the normal distribution), we run $1,000$ simulations each for wild-type, \emph{nacre}, \emph{pfeffer}, and \emph{shady}. \footnote{When we add noise to the annulus parameters, we choose both the inner radius and the annulus width from a normal distribution.} Our goal in this study is two-fold: first, we aim to make quantitative predictions comparing variability in wild-type and mutant patterns, and second, we seek to identify the range of patterns these fish may display in the presence of stochastic cellular communication.

In order to quantitatively explore how additional stochasticity impacts patterning, we first need to define what it means for a pattern to look the same as (or different from) what we would expect characteristically. For wild-type, this is immediate: we characterize wild-type patterns in terms of stripe and interstripe breaks. For 
\emph{nacre}, \emph{pfeffer}, and \emph{shady}, however, the process is more challenging because these mutant patterns are messier. For example, from looking at the images of \emph{nacre} in Fig.~\ref{fig:bio}B and Fig.~\ref{fig:bio}G, it is not clear at what point \emph{in silico} patterns consisting of elongated, orange globs should be considered good or bad matches for \emph{nacre}. This is where our baseline analysis of 
\emph{nacre}, \emph{pfeffer}, and \emph{shady}
plays a role. We use our 
earlier analysis of simulations in the default parameter regime to identify patterns that fall outside of what constitutes ``characteristic'' for each of these mutants (in terms of number and size of spots). 
For each mutant, we set our thresholds for small and large spots to be the minimum and maximum values, respectively, of the cluster-size measures that we found in our baseline experiments with that mutant. 
Analogously, for each mutant, we set the threshold for what constitutes few (many) spots to be the minimum (maximum) number of spots we found in our baseline simulations with that mutant.

\begin{figure*}[t!]
\centering
\includegraphics[width=\textwidth]{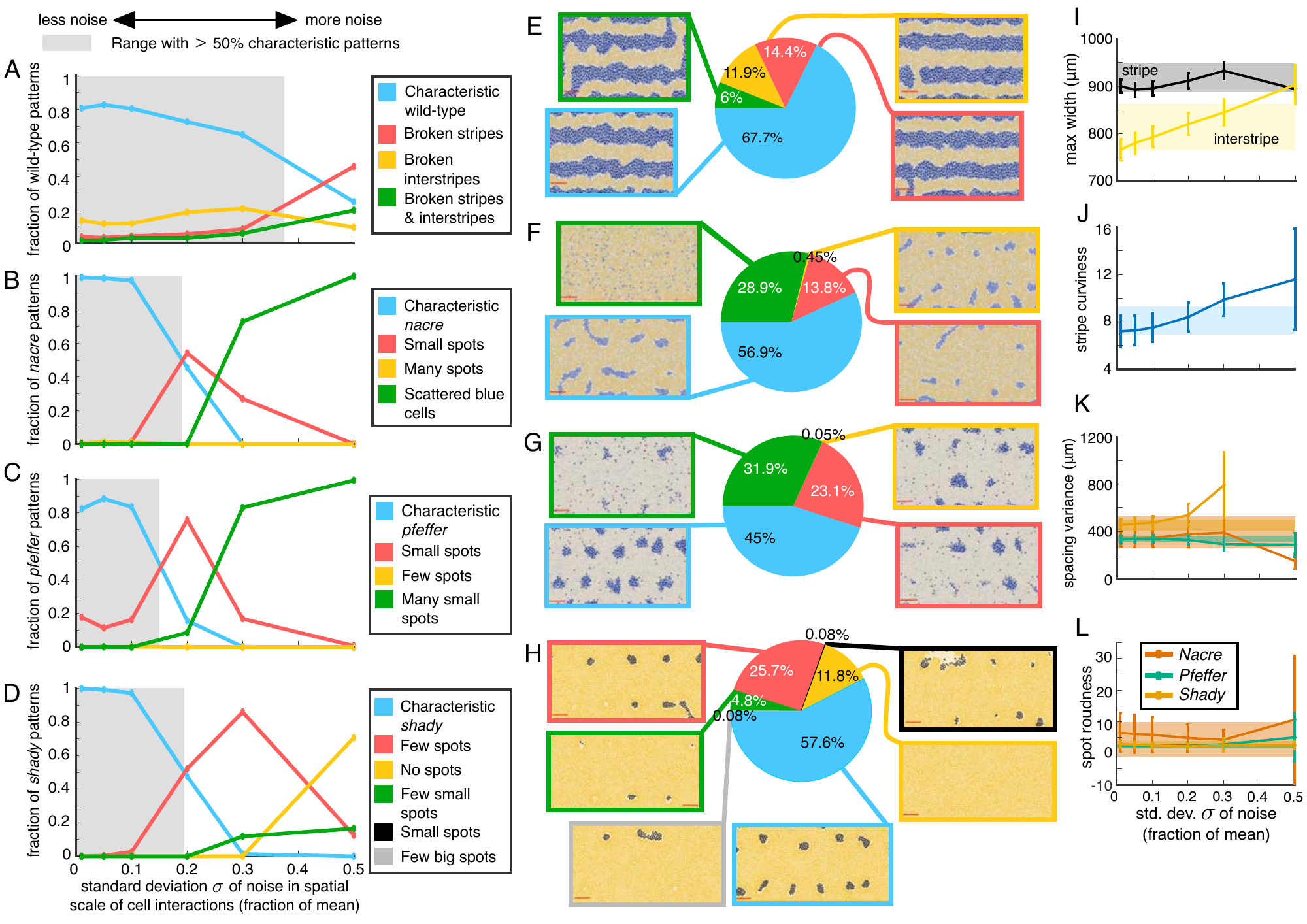}
    \caption{Quantitative study of how stochasticity in cell interactions affects wild-type and mutant zebrafish patterns. 
For each value of $\sigma \in \{0.01,0.05,0.1,0.2,0.3,0.5\}$, where $\sigma$ times the default length scale is the standard deviation of the noise that we include in the size of cellular interaction neighborhoods, we analyze $1,000$ simulations for wild-type and each mutant. Summary of the patterns that emerge under stochasticity, as detected using our methods for (A, E) wild-type, (B, F) \emph{nacre}, (C, G) \emph{pfeffer}, and (D, H) \emph{shady}. In (A--D), we highlight the range of $\sigma$ values that retain at least $50$\% characteristic patterns under noise in grey. (We define ``characteristic'' for wild-type as patterns having $3$ unbroken interstripes and $2$ unbroken stripes, and we define ``characteristic'' for mutants as patterns with spot size and spot number that fall within the baseline distributions in Fig.~\ref{fig:default_ESM_hists}A--B.) 
(I) Mean maximum stripe/interstripe width and (J) mean stripe curviness for wild-type for different noise strengths. (K) Spot spacing variance and (L) spot roundness for mutants under different noise strengths. In (I--L), the bars indicate standard deviation and the shaded regions give the characteristic values (the mean plus/minus one standard deviation) for the associated measurements  from our default studies.
Numbers in (E--H) are given to three significant figures; also see Tables S2--S5.}
\label{fig:pcpd_donut}
\end{figure*}

In Fig.~\ref{fig:pcpd_donut}A--D, we show how prevalent 
various patterns are across our stochastic simulations for different levels of noise in cell-interaction length scales (see Tables S2--S5 in the Supplementary Materials for additional measurements).
As an agglomerate summary across all $6,000$ simulations that we generated for different $\sigma$ values, Fig.~\ref{fig:pcpd_donut}E--H provides examples of the different patterns categorized by our methods for wild-type and each mutant. Our results in Fig.~\ref{fig:pcpd_donut}A--D suggest that wild-type and mutant patterns behave differently in the presence of noise. In particular, all three mutants have characteristic spots in less than 50\% of the model outputs when $\sigma \geq 0.2$, while wild-type patterns retain characteristic unbroken stripes and interstripes more robustly. If we take a closer look at individual pattern features in Fig.~\ref{fig:pcpd_donut}I--J, we note that low levels of noise ($\sigma \le 0.1$) 
serve to straighten stripes and that stripe width is mostly unaffected by the inclusion of noise in cell size and projection length. As stochasticity increases, wild-type patterns display a gradual decay in quality over the range of $\sigma$ values that we consider. With increasing noise, we find more breaks in interstripes, wider interstripes, curvier stripes, and marginally slower pattern formation (see Table S2). Wild-type stripes do not appear to completely deviate from 
characteristic until $\sigma = 0.5$, at which point broken stripes become the norm.

In comparison, the mutant patterns are almost unaffected by noise for $\sigma \le 0.1$, but then undergo a sharp change in pattern features as $\sigma$ increases. When \emph{nacre} and \emph{pfeffer} stray from characteristic, we mostly observe small spots or scattered cells (Fig.~\ref{fig:pcpd_donut}B--C). Noisy length scales in \emph{shady}, in turn, generally produce patterns with few or no dark spots (Fig.~\ref{fig:pcpd_donut}D). Related, Frohnh\"{o}fer \emph{et al.} \cite{Frohnhofer} observed that strong forms of the \emph{shady} mutant have no spots. As we show in Fig.~\ref{fig:pcpd_donut}B--D and Tables S3--S5, spots on all three mutants retain their characteristic roundness across a range of $\sigma$ values, only deviating substantially from the measures in Fig.~\ref{fig:default_ESM_hists} when $\sigma = 0.5$.

To roughly approximate the amount of noise present in cellular length scales \emph{in vivo}, we estimate the standard deviation reported for the distance between neighboring xanthophores \cite{2016heterotypic} and the length of their filopodia extensions \cite{Mahalwar}. 
Based on graphs in \cite{2016heterotypic}, we estimate that the distance between the centers of neighboring $X^\text{d}$ cells
(at $40$ dpf) is $27$~$\mu$m with a standard deviation of $4.6$ $\mu$m; in our notation, this means that $\sigma = 27/4.6$, so the standard deviation is about $17$\% of the mean. Similarly, using graphs in the SI of \cite{Mahalwar}, we estimate that the longest xanthophore 
extensions (measured from the cell center) have a standard deviation in length that corresponds to $12$\% and $20$\% of the mean filopodia lengths before and after iridophores arrive on the skin, respectively (in particular, we find that the filopodia length before iridophores arrive is approximately $58$ $\mu$m $\pm$ $6.7$ $\mu$m, and the filopodia length after iridophores arrive is approximately $25$~$\mu$m~$\pm$~$5$~$\mu$m). These measurements suggest that focusing on the patterns that emerge when $\sigma$ is between roughly $0.1$ and $0.2$ in our simulations may have particular biological relevance. We caution that this approximation is based on variance in short-range length scales only, and cells may also communicate through long-range projections \cite{delta,eom2015long} (as well as diffusion of signaling molecules \cite{PatNcomm}); moreover, in comparing these measurements to our simulations, we are inherently assuming that the empirical data has a normal distribution.

 Motivated by our 
 estimates of standard deviation \emph{in vivo},
 we explore what our analysis predicts when $\sigma \in [0.1,0.2]$.   As we note in Table S2, we find that wild-type stripe width, stripe curviness, and the time of formation of interstripes X1V and X1D  are robust in this range of $\sigma$. Our 
 methods allow us to estimate that $84.8$\% and $78.1$\% of the wild-type patterns for $\sigma =0.1$ and $\sigma =0.2$, respectively, feature characteristic unbroken interstripes (recall, $87.5$\% of our simulations in the baseline experiments with $\sigma = 0$ have unbroken interstripes). Echoing empirical observations \cite{Frohnhofer} that mutant patterns are more variable than wild-type, we find that the model \cite{volkening2} supports a distribution of mutant patterns for $\sigma \in [0.1,0.2]$. In particular, we predict that the representative images of \emph{nacre}, \emph{pfeffer}, and \emph{shady} in Fig.~\ref{fig:bio}B--D and Fig.~\ref{fig:bio}F--H are characteristic of these mutants in the sense that roughly half of the associated fish may
 resemble them, while we expect that the remaining fish resemble versions of these images with fewer and smaller spots. 
 Crucially, we predict that the mutants do not commonly display larger spots than those in Fig.~\ref{fig:bio}F--H. In the future, analyzing extensive collections of empirical images
 will allow one to test our predictions and the model \cite{volkening2}.

\subsection*{A Means of Linking Altered Cell Behavior to Mutant Patterns\label{sec:LSTest}}

Thus far, we have focused on exploring wild-type patterns and the \emph{nacre}, \emph{pfeffer}, and \emph{shady} mutants. Based on transplantation experiments \cite{Maderspacher2003,ParTur130,Frohnhofer}, these mutant patterns seem to arise 
because a cell type is missing, rather than due to altered cell interactions. Zebrafish also feature a second type of mutant pattern that forms because cell behavior is altered (often in unknown ways) despite all cell types being present. Examples of this second type of mutant include \emph{leopard} and \emph{obelix},
 which feature spots and widened stripes, respectively \cite{Maderspacher2003}. Mutations that alter cell behavior provide modelers with an opportunity to help link genes to cellular function. (We note that many zebrafish genes have an orthologue in the human genome \cite{genome}.) One can adjust cell behavior in a model to search for patterns that match various mutants; in this way, a modeling approach can help establish
links between cell behaviors and the genes that control them through the phenotype. Agent-based models (e.g., \cite{volkening, volkening2,Shinbrot}) often have a large number of parameters, however, and this makes it challenging to comprehensively screen for the cellular interactions that may be related to various mutants by adjusting parameters and visually inspecting the resulting simulations. In a similar vein, modelers seek to
present a broad picture of the impact of varying different parameters, but this process is again often limited by the time-consuming nature of visual inspection. We expect that our 
methods can be used to help address these challenges, and we provide one example to illustrate this process next.

As an example study, we vary a single parameter in the model \cite{volkening2} across a range of values and apply our methods to the resulting patterns. In particular, we focus on the cellular interaction radius represented by $\Omega_\text{long}$ in \eqref{rule:Mbirth}. As shown in Fig.~\ref{fig:bio}J and discussed in \textit{Background and Methods}, long-range 
interactions depend on the proportion of cells in an annulus region $\Omega_\text{long}$ in the model \cite{volkening2}. \eqref{rule:Mbirth} describes $M$ birth as occurring at randomly selected locations $\textbf{z}$ when the number of $I^\text{d}$ and $X^\text{d}$ cells in 
$\Omega^\textbf{z}_\text{long}$ is sufficiently larger than the number of $M$ in this annulus. This models empirical observations that $M$ differentiate from precursors or stem cells \cite{Budi,Dooley,Singh} and that $X^\text{d}$ and $I^\text{d}$ in neighboring interstripes support black cell birth, while other $M$ inhibit it \cite{Nakamasu,Patterson2013} at long-range. In \cite{volkening2}, the inner radius of the annulus $\Omega_\text{long}$ is $210$ $\mu$m (motivated by \emph{in vivo} measurements of cellular extensions in \cite{eom2015long,delta}) and the width of the annulus is $40$ $\mu$m. Here we vary the inner radius parameter from $10$ to $400$ $\mu$m in increments of $25$ $\mu$m and run $100$ simulations 
under each parameter regime for wild-type, \emph{pfeffer}, and \emph{shady}\footnote{Note that we do no simulate \emph{nacre} because this mutant has no $M$ cells.}. This allows us to comprehensively explore the impact of long-range signaling on $M$ differentiation.

\begin{figure*}[t!]
    \centering
    \includegraphics[width=\textwidth]{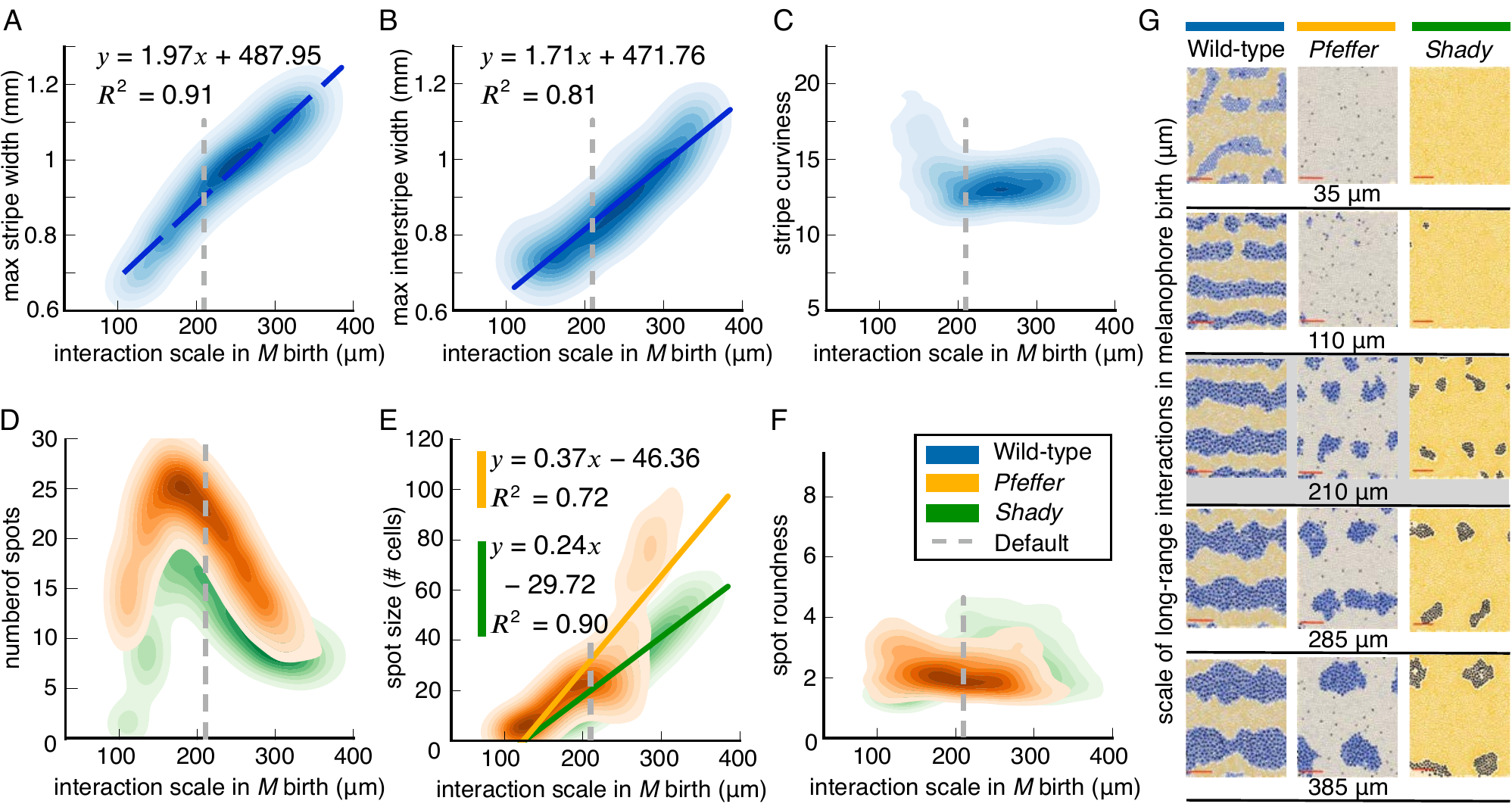}
    \caption{Quantifying \emph{in silico} pattern dependence on the spatial scale of long-range cellular interactions involved in $M$ birth. Kernel density estimates for (A--B) maximum stripe and interstripe width for wild-type, (C) wild-type stripe curviness, (D) number of spots for \emph{pfeffer} and \emph{shady}, (E) median spot size for the mutants, and (F) \emph{pfeffer} and \emph{shady} spot roundness as a function of the inner radius of the $\Omega_\text{long}$ neighborhood in \eqref{rule:Mbirth}. Measurements in (A--F) are based on $100$ simulations of the model \cite{volkening2} (for wild-type, \emph{pfeffer}, and \emph{nacre}, respectively) for each inner radius $R$ of $\Omega_\text{long}$ in \eqref{rule:Mbirth} considered.
    (We consider $R$ from $10$ to $400$ $\mu$m in increments of $25$ $\mu$m.) 
    All other model parameters (including the width of the $\Omega_\text{long}$ annulus in \eqref{rule:Mbirth} and the long-range annulus scale in all other model rules) remain at their default values. In (A), (B), and (E) we show linear regression models for their corresponding values, along with the $R^2$ goodness-of-fit scores. (G) Example wild-type, \emph{pfeffer} and \emph{shady} patterns for different parameter values (the patterns generated by the model \cite{volkening2} under the default parameter -- $210$ $\mu$m -- are noted in gray).    }
    \label{fig:MBA_kde}
\end{figure*}

If $\Omega_\text{long}$ in \eqref{rule:Mbirth} is too small (e.g., when its inner radius is below $30$--$80$ $\mu$m, the average distance between 
cells \cite{TakahashiMelDisperse,2016heterotypic}), it is likely that there are
no or very few cells in this annulus region, so that the signal from $X^\text{d}$ and $I^\text{d}$ to promote $M$ cell birth is effectively turned off.
Intuitively, this should lead to an $M$ shortage in the resulting patterns. 
Conversely, we expect that increasing the inner radius of $\Omega_\text{long}$ will widen black stripes.
To test these hypotheses and determine the role of this parameter in wild-type, we use our methodology to measure 
stripe width, interstripe width, and stripe curviness across a range of $\Omega_\text{long}$ values. For \emph{pfeffer} and \emph{shady}, we compute spot size, number, and roundness as a function of $\Omega_\text{long}$ in \eqref{rule:Mbirth}. We present our results in Fig.~\ref{fig:MBA_kde} using kernel density estimation plots to visualize the $2$--D probability density function of pattern features and parameter values. 

As we show in Fig.~\ref{fig:MBA_kde}A--B and \ref{fig:MBA_kde}E, there is a strong positive correlation between the spatial scale of long-range signaling in $M$ birth and stripe width, interstripe width, and spot size. To check that the quantities we detected automatically agree with results by visual inspection, we show a few sample simulations in Fig.~\ref{fig:MBA_kde}G for different $\Omega_\text{long}$ values. As expected, we find that the width of black stripes in wild-type increases as the spatial scale of long-range signals promoting $M$ birth increases. Conversely,
when the scale of $M$-birth signals in \eqref{rule:Mbirth} is very local,
the resulting patterns vaguely resemble the
\emph{nacre} mutant, which features no melanophores \cite{Maderspacher2003,Frohnhofer}. This highlights the importance of large-scale simulations and automated 
methods, as they allow comprehensive model explorations and provide a more complete picture of the roles of different parameters.

To further explore our results,
we ran a linear regression analysis on the pattern quantities that we present in Fig.~\ref{fig:MBA_kde}.
For wild-type, we find that a linear model in stripe width yields a coefficient of determination $R^2 = 0.912$, meaning that the linear model captures $91.2\%$ of the observed stripe width variance. For \textit{shady}, a linear model in spot size has a corresponding $R^2 = 0.901$, while a linear model in spot size for \textit{pfeffer} has a lower goodness-of-fit ($R^2 = 0.722$) because the spot size increases more rapidly. Regression models of this type can be used to predict pattern quantities without needing any reference data. In particular, these simple regression models have the potential to allow one to predict pattern features as a function of cellular interaction signals without needing to perform any model simulations.

The results of our case study exploring the impact of a single parameter (related to long-range signals in melanophore differentiation) show promise. In particular, they suggest that our methods can be applied not only for pattern quantification but also for model sensitivity analysis and large-scale parameter screening to detect possible ways that cell interactions may be altered in mutations.
We note that our goal here is
to highlight another way our methods can be applied, and we leave a more thorough investigation of zebrafish mutations and the altered cell interactions involved for future work. For example, the \emph{obelix} mutant \cite{Maderspacher2003} features widened stripes due to unknown altered cell interactions; by systematically varying parameters in the model \cite{volkening2} and automatically detecting their impact on stripe width, one could identify a set of altered cell behaviors that may be responsible for this phenotype, and these predictions could then be evaluated experimentally.

\section*{Discussion and Conclusions }

Our goal was to provide methods for quantifying 
agent-based patterns across a range of 
scales. Leveraging topological data analysis and machine learning, we developed a new methodology that captures information spanning local features of interacting cells up to macroscopic spots and stripes. Because it describes shape features across a sequence of spatial scales, persistent homology is a critical tool in our methods. We showed that combining this topological tool with clustering methods yields a collection of summary statistics that can be automatically extracted from patterns using agent coordinates. By reducing the role of visual inspection in describing patterns, our interpretable methodology provides a new means of analyzing large data sets and studying how stochasticity in agent interactions affects pattern variability. To illustrate the promise of our methods, we applied our methodology to an extensive data set of \emph{in silico} zebrafish skin patterns that we generated using the agent-based model \cite{volkening2}. Our methods allowed us to make quantitative predictions about the types and amounts of variability that may arise in wild-type and mutant zebrafish patterns due to stochasticity in cellular communication.
We used our methods to distinguish and characterize similar mutant patterns, and we showed how to track pattern features across spatial scales to study the role of different cellular interactions in pattern formation.

Many of our results, which provide a broader view of the agent-based model \cite{volkening2}, can be experimentally tested in the future. In particular, after extracting cell coordinates from zebrafish images, one could compute summary statistics for the empirical data and compare these measurements to our simulations. Our methods could also be applied to other 
models of zebrafish patterning, including partial differential equations (e.g., \cite{Yamaguchi,Nakamasu,Bullara,Painter}, stochastic cellular automaton perspectives \cite{Bullara,MorDeutsch}, and agent-based models (e.g., \cite{volkening,Shinbrot}).  In the future, one could use our methods to optimize model parameters or conduct large screens for cell interactions that may be altered in mutations. Although we focused primarily on analyzing zebrafish patterns at a fixed point in development, future work could track pattern features across developmental timelines.

Our approach to quantifying zebrafish patterns begins to address major challenges associated with quantifying agent-based dynamics
in an objective and automated 
 way, but there are also limitations to our methods. First, we make underlying assumptions about the patterns that we are studying. As an example, when we use topological methods to quantify spots or stripes,
we assume that the input patterns have certain features (e.g., we assume a wild-type input has stripe patterns). It may be useful for future studies to automatically classify each input pattern as spots or stripes prior to applying the appropriate pattern quantification methods. Moreover, we focused primarily on spots and stripes, but methods for characterizing other patterns (e.g., labyrinth patterns on the \emph{choker} mutant \cite{Frohnhofer}) could be developed in the future. Lastly, we note that we built our methodology to take data in the form of agent coordinates. Empirical images and simulations from partial differential equation models, however, are continuous functions defined over $2$--D domains. In the former case, one option would be to extract cell locations from image data, and, in the latter, one could apply our methods to cell densities after discretizing space and applying a density threshold. Fortunately, functional persistent homology could avoid both of these extra steps as it takes function data as its input. In the future, one could apply our approach to continuous pattern data by replacing the TDA tools that we used
with functional persistence throughout our methodology.

Although we focused on analyzing pattern variability in zebrafish, we expect that a similar approach can be used to quantify agent-based dynamics in other biological settings.  Methods that provide summary statistics for pattern features across a range of length scales open up many possibilities for
quantitatively comparing large data sets of \emph{in silico} and \emph{in vivo} pattern data in the future. By working closely with the needs of each application, we expect that our topological perspective can be extended to analyze agent-based dynamics in wound healing, animal flocks, and other forms of collective behavior.

\paragraph{Data and code availability}{
The code we developed for our methodology will be made available upon publication, along with zebrafish data from our simulations. We performed all TDA computations using Ripser \cite{Ripser} and completed post-processing in MATLAB R2017b and Python 3.6.4.}

\paragraph{Author contributions}{M.R.M, A.V., and B.S. designed methodology and experiments; M.R.M. wrote code for the pattern quantification methods and performed experiments; A.V. provided biological analyses; M.R.M, and A.V. wrote the paper. All authors contributed to revisions and gave final approval of the manuscript.}

\paragraph{Competing Interests}{We declare we have no competing interests.}

\paragraph{Acknowledgements}{We thank Richard Carthew, Anastasia Eskova, Hans Georg Frohnh\"{o}fer, Uwe Irion, Marco Podobnik, and Christiane N\"{u}sslein-Volhard for helpful comments and discussion.
M.R.M.\ is supported by the National Science Foundation (NSF) Graduate Research Fellowship Program under grant no.\ 1644760.
B.S. is partially supported by the NSF under grants DMS-1714429 and CCF-1740741.
A.V.\ has been supported by the Mathematical Biosciences Institute and the NSF under grant no.\ DMS-1440386 and is currently supported by the NSF under grant no.\ DMS-1764421 and by the Simons Foundation under grant no.\ 597491.}

\section*{Supplementary Materials}
\def\theequation{S\arabic{equation}}
\def\thefigure{S\arabic{figure}}
\def\thetable{S\arabic{table}}
\setcounter{figure}{0}
\setcounter{equation}{0}
\setcounter{table}{0}

\subsection*{Additional Background: Persistent Homology}

Topological data analysis (TDA) is an emerging branch of applied mathematics that aims to extract useful shape descriptors from large, complex data sets  \cite{Carlsson2009,chazal, Edelsbrunner10, Ghrist2014, Zomorodian2009}. The utility of TDA has been demonstrated in a range of applications, including neuroscience (e.g., \cite{Curto2017, Giusti13455, Sizemore2018}), genomics (e.g., \cite{Chan2013, camera2016, Humphreys2019}, and sensor networks (e.g., \cite{Munch2012, Adams2015}). Persistent homology is one of the main techniques in TDA, and it involves the construction of simplicial-complex representations of a given data set across a sequence of scales.
We provide an introduction to persistent homology here. 

The goal of persistent homology is to associate homology groups to data. For the scope of our paper, we can think of homology groups as vector-space representations of a topological object  with dimension 
corresponding to the number of ``holes'' 
that object has. (For more details on algebraic topology, see \cite{Hatcher:478079}.) Specifically, the $0$th 
and $1$st dimension homology groups of an object are vector spaces whose dimensions are the number of connected components and loops, respectively, 
that the object has. For example, a figure-eight object has a single connected component 
and two loops (see Fig.~2 in the main manuscript and Fig.~\ref{fig:Figure8_PD}). In other words, the $0$th homology group of a figure eight has one generator, and its first homology group has two generators. 

Next consider a data set $S = \{x_i\}_{i \in I}$, where $S$ is any collection of points living in a given metric space $(D, d)$. For example, $S$ could be a set containing the coordinates of all of the pigment cells in a zebrafish skin pattern. If we were to extract the homology groups of $S$ directly, we would get uninformative representations. In particular, the $0$th homology group of $S$ has dimension equal to the number of data points in $S$ and all of the higher homology groups are trivial (e.g., if $S$ contained the coordinates of pigment cells, the $0$th homology group of $S$ would simply be the number of cells in $S$). Instead, in persistent homology, one builds simplicial-complex representations of $S$ to infer the homology of the manifold from which the data were sampled. In our example, this allows us to study the qualities of the stripes and spots that are made up of the pigment cell coordinates in $S$.

\begin{figure}[h!]
    \centering
    \includegraphics[width=\textwidth]{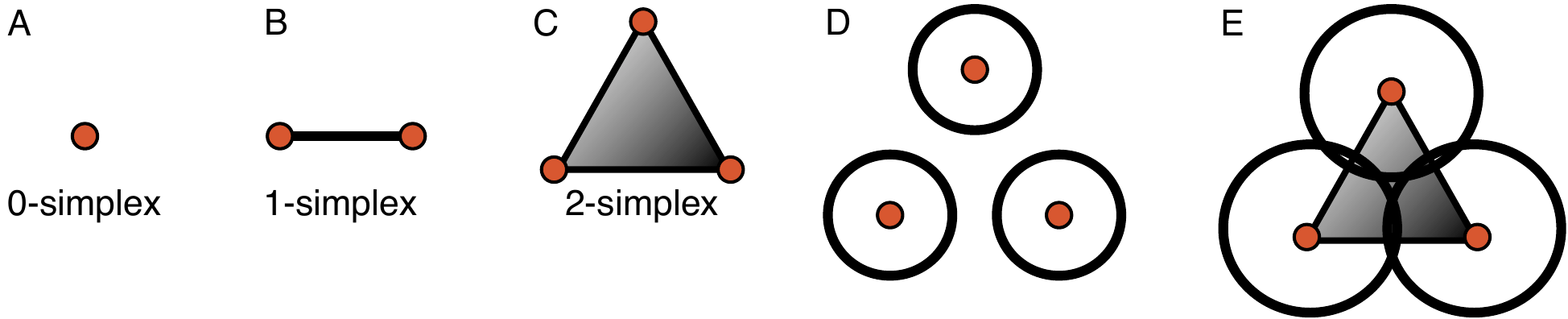}
    \caption{Introduction to simplicial complexes. Examples of (A) a $0$-simplex (point), (B) a $1$-simplex (edge or line), and (C) a $2$-simplex (triangle). (D--E) Given $3$ points (namely, $x_1$, $x_2$, and $x_3$), we show their Vietoris--Rips simplicial complex with respect to two separate radius parameters. In panel (D), the radius parameter $r$ is small enough so that $B_r(x_i) \cap B_r(x_j) = \emptyset$ for all $i,j = 1,2,3$, $i \neq j$. Consequently, the Vietoris--Rips complex with respect to the $r$ value in (D) simply consists of three $0$-simplices: $\sigma_0(x_1)$, $\sigma_0(x_2)$, and $\sigma_0(x_3)$. As the radius parameter $r$ grows in panel (E), we find pairwise non-empty intersections $B_r(x_1) \cap B_r(x_2) \neq \emptyset$, $B_r(x_1) \cap B_r(x_3) \neq \emptyset$, and $B_r(x_2) \cap B_r(x_3) \neq \emptyset$, so the resulting Vietoris--Rips simplicial complex is the union of three $0$-simplices $\sigma_0(x_1)$, $\sigma_0(x_2)$, and $\sigma_0(x_3)$; three $1$-simplices $\sigma_1(x_1, x_2)$, $\sigma_1(x_1, x_3)$, and $\sigma_1(x_2, x_3)$; and one $2$-simplex $\sigma_2(x_1, x_2, x_3)$. }
    \label{fig:simplicial}
\end{figure}

\begin{figure}[t!]
    \centering
    \includegraphics[width=\textwidth]{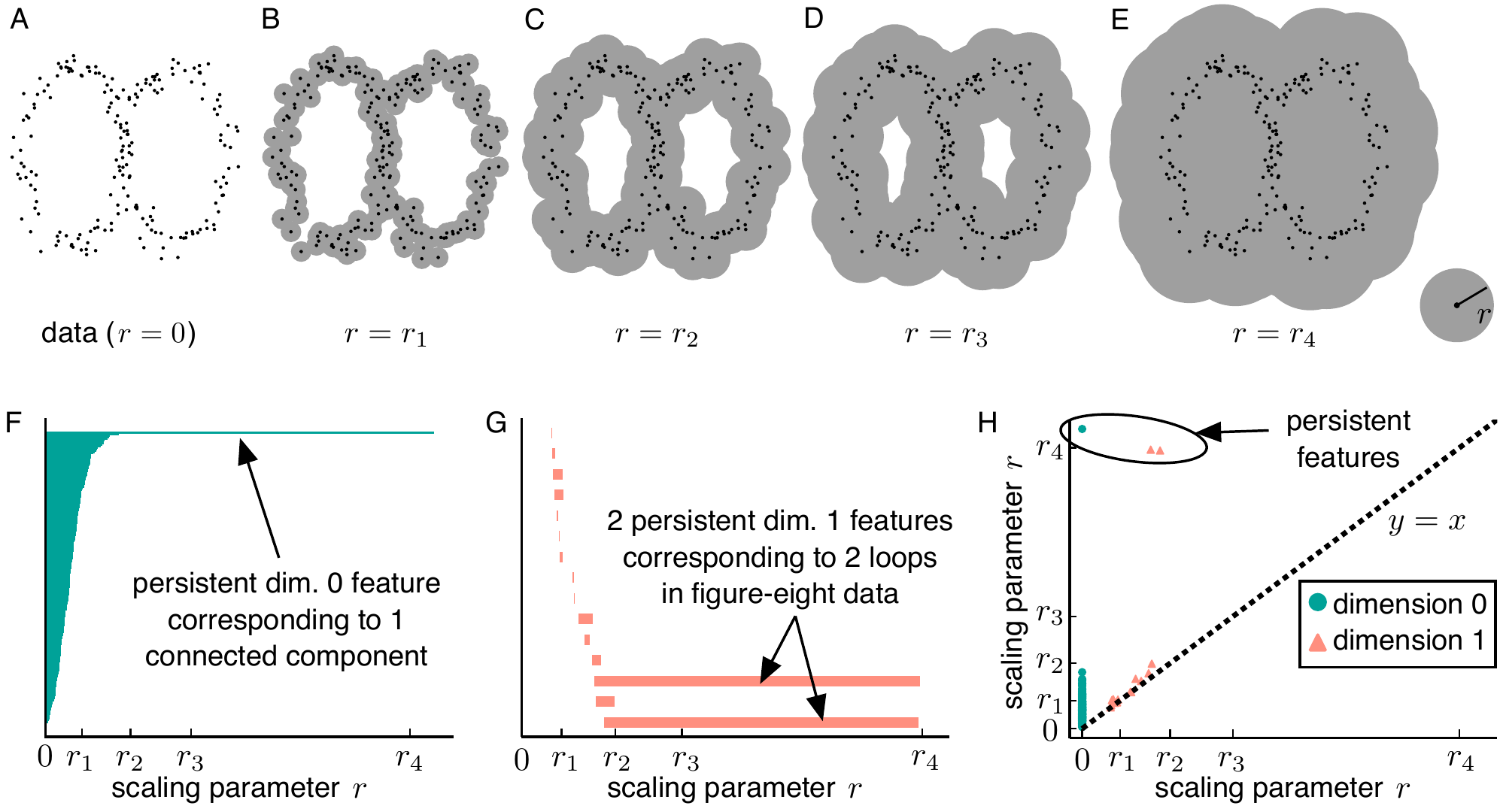}
    \caption{Illustration of persistent homology applied to coordinate data and corresponding barcode and persistence diagrams. For easier comparison with our barcode and persistence diagrams, here we show (A) original coordinate data sampled from a figure-eight shape and (B--E) corresponding manifold expansions given by $\{b_r(\textbf{x}_i)\}_{i =1}^N = \{\textbf{y} \in D, i \in [1,N]: d_D(\textbf{x}_i, \textbf{y}) \leq r\}$ for $r = r_1 < r_2 < r_3 < r_4$ (also see Fig.~2 in the main text). (F--G) We show the dimension $0$ and dimension $1$ barcode diagrams that correspond to persistent homology applied to the figure-eight data. The long bar in panel (F) represents the single connected component of the figure-eight shape, and the two long bars in (G) correspond to the two loops of the figure-eight shape. (H) As an alternative means of viewing persistent homology results, we also show the analogous persistence diagram.  The circular point in teal in the top left corner of the diagram represents the persistent dimension $0$ feature (namely, the single connected component in our figure-eight data); and, the two triangular points in pink in the top left corner of the diagram correspond to the two persistent dimension $1$ features (namely, the two loops in our figure-eight data). We calculated persistent homology using Ripser \cite{Ripser}. 
    }
     \label{fig:Figure8_PD}
\end{figure}

A $k$-simplex of $k + 1$ affinely independent points is the convex polygon whose vertices are precisely the $k + 1$ affinely independent points \cite{Ghrist2014, Edelsbrunner10}. In other words, a $0$-simplex $\sigma_0(x_1)$ is a single point $x_1$; a $1$-simplex $\sigma_1(x_1, x_2)$ is an edge connecting $x_1$ and $x_2$; a $2$-simplex $\sigma_2(x_1,x_2, x_3)$ is a filled triangle whose vertices are $x_1,x_2$, and $x_3$; and so on. To begin building 
simplicial-complex representations of $S$, we 
place a ball of radius $r$ centered 
at each $x_i \in S$ to obtain $\{ B_{r}(x_i) = \{y \in D : d(x_i, z) \leq r \} \}_{i \in I}$. The simplicial complex representation of $S$ with respect to $r$ is the union of all k-simplices $\sigma_k(x_{j_0}, \ldots, x_{j_k})$ such that $B_{r}(x_{j_l}) \cap B_{r}(x_{j_j}) \neq \emptyset$ for all $l, j = 0, 1, \ldots, k$. This is known as the Vietoris--Rips complex of $S$ with respect to $r$. We can now compute the homology groups of this simplicial-complex representation of $S$. See Fig.~\ref{fig:simplicial} for examples of different simplicies.

Inevitably, the homology groups of the simplicial-complex representation of $S$ will be sensitive to the choice of scaling parameter $r$. To overcome this challenge, persistent homology instead tracks how the homology groups change across an increasing sequence of scaling parameters $\{r_j \}_{j \in J}$. The scaling parameter at which a homological generator appears is called the birth radius ($r_b$) of a topological feature. Similarly, the 
value of $r$ at which a homological generator disappears is called the death radius ($r_d$) of a topological feature. The persistence of a topological feature is its lifetime, namely $r_d -r_b$.

In TDA there are two primary ways of visualizing the persistent homology of a data set. The first method, called a barcode diagram, is a collection of bars. Each bar in a barcode diagram represents a topological generator; the left endpoint of a bar corresponds to the birth radius of that generator, and the right endpoint of a bar corresponds to its death radius. Topological features that persist for across a longer range of radii values
are identified by having longer bar representations in the barcode diagram. The second method, called a persistence diagram, is a collection of points in $\mathbb{R}^2_{\geq 0}$. Each point in a persistence diagram corresponds to a topological feature, and the $x$- and $y$-coordinates of the points are the birth and death radii, respectively, of those features. In a barcode diagram, topological features that persist across a broader range of parameter values $r$ lie farthest away from the diagonal line $y = x$. These persistent homology visualizations provide a low-dimensional, descriptive representation of the original data $S$ and are therefore useful for a range of data-driven tasks. For example, in Fig.~\ref{fig:Figure8_PD}F--G we show the dimension $0$ and dimension $1$ barcode diagrams corresponding to the figure-eight example in the main text (see Fig.~2 in the main text). We show the analogous persistence diagram for this example in Fig.~\ref{fig:Figure8_PD}H.  The long bar in Fig.~\ref{fig:Figure8_PD}F represents the single connected component of the figure-eight shape and the two long bars in Fig.~\ref{fig:Figure8_PD}G correspond to the two loops of the figure-eight shape. In the persistence diagram, these topological features are represented by the top left circular teal point in Fig.~\ref{fig:Figure8_PD}H (dimension $0$ feature) and the two triangular pink points in the top left corner in Fig.~\ref{fig:Figure8_PD}H (dimension $1$ features).

\subsection*{Measuring Local Pattern Features\label{sec:tools-LAF}}

While TDA and machine learning offer exciting insights into biological patterns, raw calculations of agent--agent distances and agent-density counts are still important measurements. We therefore use direct calculations in tandem with the aforementioned methods for tracking pattern variability. In particular, we calculate the mean and variance of nearest-neighbor distances between agents directly using location data. The coefficient of variation (CV) for the distance between neighboring agents provides an additional measurement of pattern quality: 
 \begin{align} \text{CV} = 100 \times \frac{\text{std} (\text{agent--agent distances})}{\text{mean}( \text{agent--agent distances})}. \label{eq:CV}
 \end{align}
 
 \begin{figure}[h!]
    \centering
    \includegraphics[width=\textwidth]{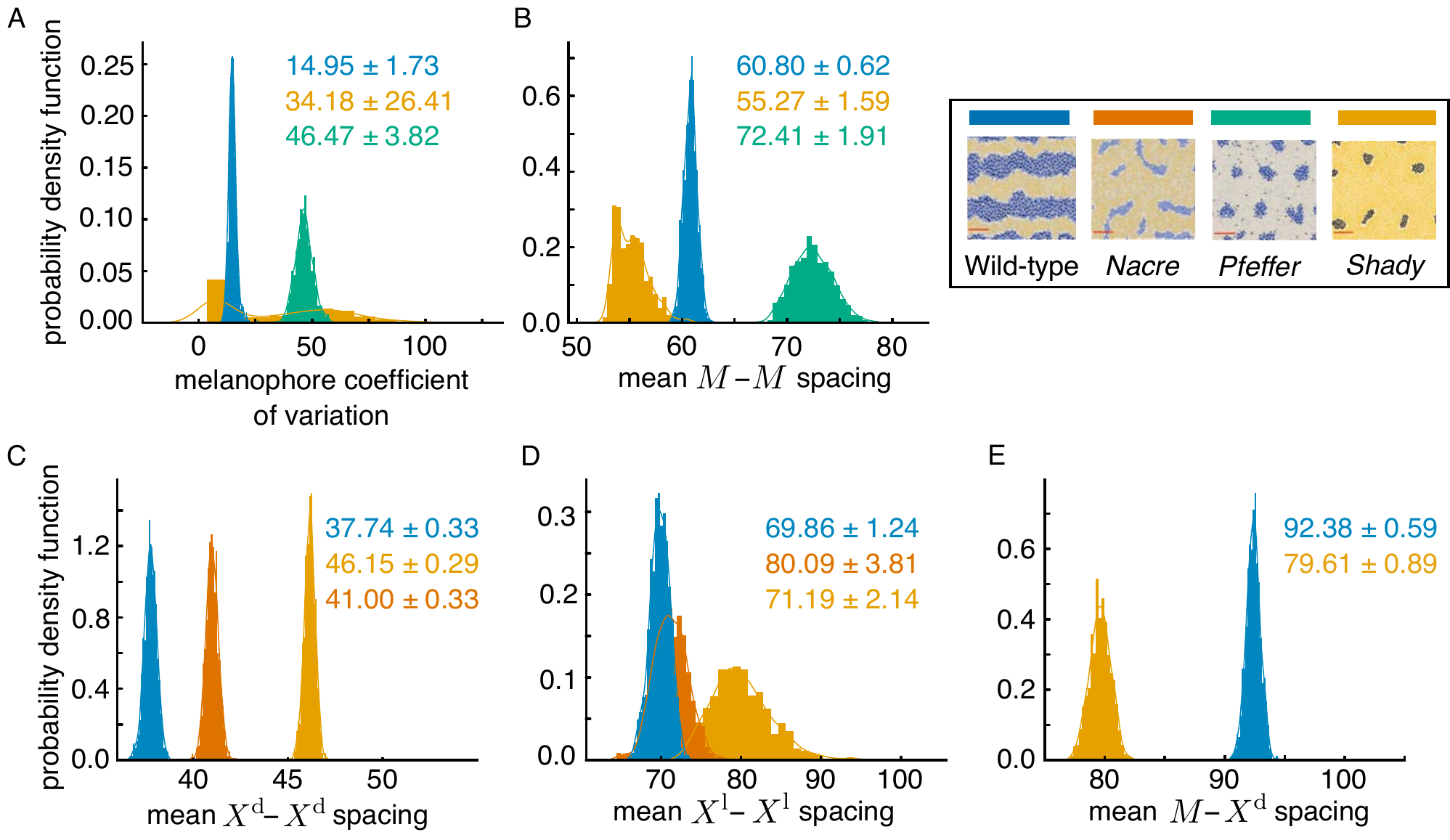}
    \caption{Measurements of local pattern features. We show distribution plots for (A) melanophore coefficient of variation \cite{ParTur130,ParTur256}, (B) mean $M$--$M$ spacing, (C) mean $X^\text{d}$--$X^\text{d}$ spacing, (D) mean $X^\text{l}$--$X^\text{l}$ spacing, and (E) mean $M$--$X^\text{d}$ spacing. We base these distributions for wild-type, \emph{nacre}, \emph{pfeffer}, and \emph{shady} patterns on $1,000$ model simulations (for each pattern type) under the default parameter regime in \cite{volkening2}. We indicate the mean plus/minus the standard deviation for these measurements in each figure. }
    \label{fig:default_spacing_hists}
\end{figure}

For zebrafish, low values of the $M$ CV are associated with better-formed patterns \cite{ParTur256,ParTur130}. 
We compute $A$--$B$ agent density counts by counting the number of agents of type $A$ per local neighborhood of the query agent type $B$, where local neighborhoods are defined as disks of radius $R$ centered at the query agent (we use $R = 250$ $\mu m$ for $X^\text{l}$ cells and $R = 200$ $\mu m$ for $I^\text{l}$ cells).
We then use the $20$th percentile of agent counts across all local neighborhoods as our summary statistic (other summary statistics could also be used). 
It is helpful to note that the $A$--$A$ agent density is just the traditional measurement of the density of agent $A$, but we find that extending this definition to allow for counting one agent type against another is helpful in some cases. For example, in the \emph{pfeffer} mutant pattern (see Fig.~1C and 1G in the main manuscript), blue $I^\text{l}$ cells are present only when there are sufficient black $M$ nearby. To calculate the $I^\text{l}$--$M$ density, we count the number of $M$ cells within a local neighborhood of each $I^\text{l}$ cell and take a summary measure across all neighborhoods.

\begin{figure}[h!]
    \centering
    \includegraphics[width=\textwidth]{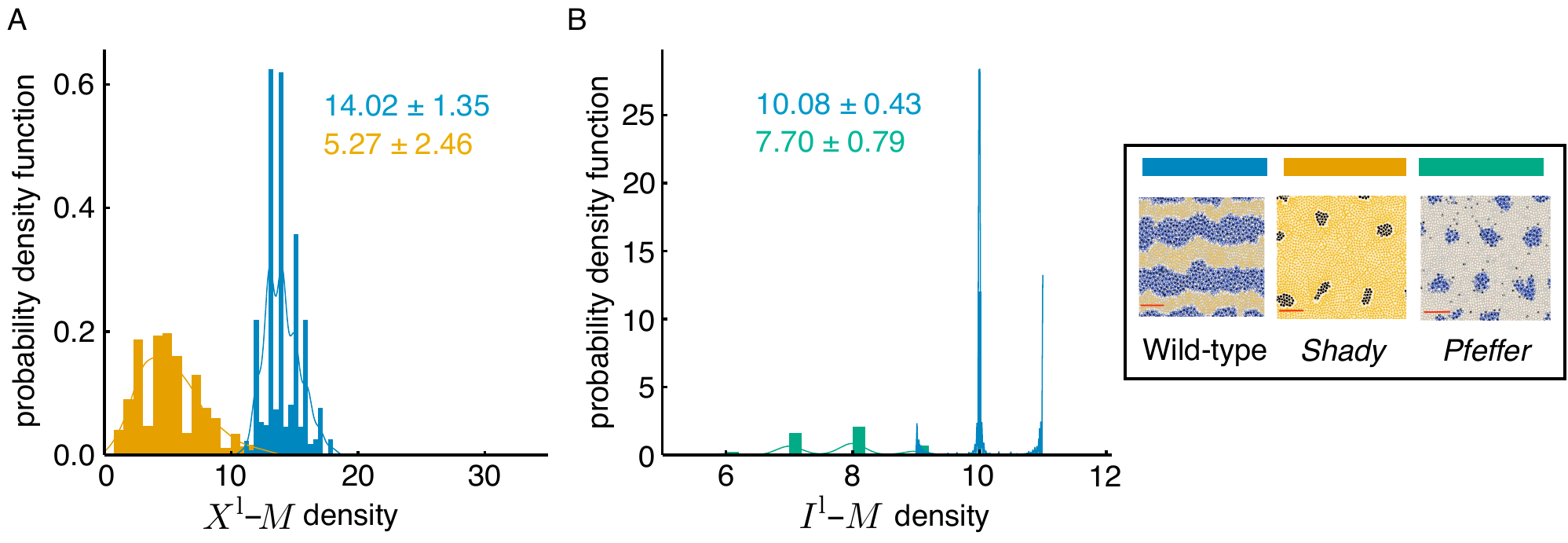}
    \caption{Measurements of cell density. We show distribution plots for (A) $X^\text{l}$--$M$ and (B) $I^\text{l}$--$M$ density counts. We base these distributions for wild-type, \emph{pfeffer}, and \emph{shady} patterns on $1,000$ simulations of the model \cite{volkening2} under the default parameter regime. (We do not show distributions for \emph{nacre} because this mutant lacks melanophores.) We indicate the mean plus/minus the standard deviation of these cell--cell density measurements in each figure.}
    \label{fig:default_density_hists}
\end{figure}

We present the $M$ CV and nearest-neighbor cell--cell spacing measurements from $1,000$ model simulations under the default parameter regime for wild-type, \nacre, \pfeffer, and \textit{shady} in Fig.~\ref{fig:default_spacing_hists}, and we show $X^\text{l}$--$M$ and $I^\text{l}$--$M$ cell-density counts in Fig.~\ref{fig:default_density_hists}. We observe that the \textit{shady} mutant has the largest variability in $M$ CV (Fig.~\ref{fig:default_spacing_hists}A). Moreover, the average $M$--$M$ nearest-neighbor spacing is smallest for \textit{shady} and greatest for \textit{pfeffer} (Fig.~\ref{fig:default_spacing_hists}B). Interestingly, the distributions of the mean $X^\text{d}$--$X^\text{d}$ nearest-neighbor spacing are completely disjoint for wild-type, \nacre, and \shady. The wild-type $X^\text{d}$--$X^\text{d}$ nearest-neighbor spacing is the smallest and the  \textit{shady} $X^\text{d}$--$X^\text{d}$ nearest-neighbor spacing is the largest, with all three types having low variance (Fig.~\ref{fig:default_spacing_hists}C). In contrast, the wild-type $M$--$X^\text{d}$ nearest-neighbor spacing is significantly greater than the \textit{shady}  $M$--$X^\text{d}$ nearest-neighbor spacing (Fig.~\ref{fig:default_spacing_hists}E). Finally, we note that the $X^\text{l}$--$M$ (Fig.~\ref{fig:default_density_hists}A) and $I^\text{l}$--$M$ (Fig.~\ref{fig:default_density_hists}B) density counts are greater for wild-type than they are for the mutants.

\subsection*{Summary of Model Rules Adjusted in our Study of Pattern Variability}

In Fig.~6 in the main text, we quantify the pattern variability induced by replacing the original deterministic length scales used in the model \cite{volkening2} with stochastic length scales. In particular, the model rules in \cite{volkening2} depend on
five interaction neighborhoods:  $B^\textbf{z}_{75}$, $B^\textbf{z}_{\Delta_\text{xm}}$, $B^\textbf{z}_{90}$, and $B^\textbf{z}_{90/2}$ denote the disks of radii $75$ $\mu$m, $90$ $\mu$m, $\Delta_\text{xm} = 82$ $\mu$m, 
and $90/2 = 45$ $\mu$m, respectively, centered at the position $\textbf{z}$ of the cell (or precursor) of interest, and $\Omega^\textbf{z}_\text{long}$ is the annulus of inner radius $210$ $\mu$m and width $40$ $\mu$m centered at position $\textbf{z}$. With the exception of $B^\textbf{z}_{\Delta_\text{xm}}$, which is used in the rules that prevent cell overcrowding due to continual birth, we replace these deterministic neighborhoods with stochastic versions to generate our results in Fig.~6 in the main text, as follows:
\begin{align} 
B^\textbf{z}_{75,\sigma} &= \text{disk centered at \textbf{z} with radius $ r \sim \mathcal{N}(75, \sigma \cdot 75)$ $\mu$m;} \label{eq:ls1} \\
B^\textbf{z}_{90,\sigma} &= \text{disk centered at \textbf{z} with radius $ r \sim \mathcal{N}(90, \sigma \cdot 90)$ $\mu$m;} \\
B^\textbf{z}_{90/2,\sigma} &= \text{disk centered at \textbf{z} with radius $ r/2$, $r \sim \mathcal{N}(90, \sigma \cdot{90})$ $\mu$m; and} \\
\Omega^\textbf{z}_{\text{long},\sigma} &= \text{annulus centered at \textbf{z} with inner radius $ r \sim \mathcal{N}(210, \sigma \cdot 210)$ $\mu$m and width $w \sim \mathcal{N}(40, \sigma \cdot 40)$ $\mu$m}, \label{eq:ls5}
\end{align}
where we consider $\sigma \in \{0.01,0.05,0.1,0.2,0.3,0.5\}$, so that the standard deviation of the noise is a fraction of its mean.\footnote{Note that $\sigma = 0$ corresponds to the deterministic length scales originally used in the model \cite{volkening2}.} We write $B^\textbf{z}_{90/2,\sigma}$ instead of $B^\textbf{z}_{45,\sigma}$ to stress that, when we randomly select $r \sim \mathcal{N}(90, \sigma \cdot{90})$ $\mu$m, we use that same $r$ value to define both the randomly-chosen radius of the disk $B^\textbf{z}_{90,\sigma}$ and the randomly-chosen diameter of the the disk $B^\textbf{z}_{90/2,\sigma}$. This means that the radius of $B^\textbf{z}_{90/2,\sigma}$ is always half the radius of $B^\textbf{z}_{90/2,\sigma}$ for the cell at position $\textbf{z}$ (we made this choice because the size of these disks depends on a single parameter in the model \cite{volkening2}). In all other cases, following the approach in \cite{volkening2}, the length scales involved in the disk and annulus neighborhoods are chosen independently.

 We include these stochastic cell-interaction neighborhoods in the model rules for $M$ birth, $M$ death, and cell-form transitions. Because it makes the model more amenable to testing a wide range of length scales, we also reframe the model rules in \cite{volkening2} so that the rules for $M$ birth, $M$ death, and xanthophore form transitions depend on the ratios of cell counts in the interaction neighborhoods rather than absolute cell counts, and the rules for iridophore form transitions depend on density counts. While ideally all rules would be modified to be in terms of density counts, this is only feasible for the iridophore form transition rules that have 
 only one interaction neighborhood per inequality. 
 
 We present the adjusted model rules that we use to generate our results in Fig.~6 in the main text in  \eqref{eq:mbirth}--\eqref{eq:ichange} below. We refer to \cite{volkening2} for their biological motivation; for each rule, we select each stochastic length scale  (namely, \eqref{eq:ls1}--\eqref{eq:ls5}) randomly per cell per day (i.e., the time step for cell interactions used in \cite{volkening2}). In all other cases, the parameters have the default values given in \cite{volkening2}. Specifically, we adapt the model rule for $M$ birth at position $\textbf{z}$ from \cite{volkening2} to the following:
\begin{align} &\underbrace{\frac{\sum_{i=1}^{N^\text{d}_\text{X}}{ \mathds{1}}_{\Omega^{\textbf{z}}_{\text{long},\sigma}} (\textbf{X}^\text{d}_i) + \sum_{i=1}^{N^\text{d}_\text{I}}{ \mathds{1}}_{\Omega^{\textbf{z}}_{\text{long},\sigma}} (\textbf{I}^\text{d}_i)}{\alpha + \beta \sum_{i=1}^{N_\text{M}}{ \mathds{1}}_{\Omega^{\textbf{z}}_{\text{long},\sigma}}(\textbf{M}_i)}>1}_\text{long-range signals for melanophore birth} \hskip0.6cm \text{and} \nonumber \\ 
& \underbrace{\sum_{i=1}^{N^\text{d}_\text{X}}{ \mathds{1}}_{B^{\textbf{z}}_{\Delta_{\text{xm}}}}( \textbf{X}^\text{d}_i) + \sum_{i=1}^{N_\text{M}}{ \mathds{1}}_{B^{\textbf{z}}_{\Delta_{\text{xm}}}}(\textbf{M}_i) + \sum_{i=1}^{N^\text{d}_\text{I}}{ \mathds{1}}_{B^{\textbf{z}}_{\Delta_\text{xm}}}(\textbf{I}^\text{d}_i)  \le \eta}_{\text{limiting condition to prevent overcrowding}} \hskip0.6cm \Longrightarrow \hskip0.6cm \text{melanophore birth at $\textbf{z}$,} \label{eq:mbirth}
\end{align}
where $\mathds{1}_R(\textbf{x})$ with $R \in \{B^\textbf{z}_{90/2,\sigma},B^\textbf{z}_{75,\sigma},B^\textbf{z}_{\Delta_\text{xm},\sigma},B^\textbf{z}_{90,\sigma},\Omega^\textbf{z}_\text{long}\}$  is the indicator function for the region $R$ (in particular, $\mathds{1}_R(\textbf{x}) = 1$ if $\textbf{x}$ is in the region $R$ and $0$ otherwise). Our adapted rule for melanophore death due to local competition \cite{Nakamasu,volkening2}, in turn, is given by:
\begin{align}\frac{\sum_{j=1}^{N^\text{d}_\text{X}}{ \mathds{1}}_{B^{\textbf{M}_j}_{90,\sigma}} (\textbf{X}^\text{d}_j )}{\sum_{j=1}^{N_\text{M}}{ \mathds{1}}_{B^{\textbf{M}_i}_{90,\sigma}}(\textbf{M}_i )}> \mu ~&\Longrightarrow ~\text{death of melanophore at $\textbf{M}_i$ due to local competition with $X^\text{d}$}.
\end{align}

The model \cite{volkening2} also includes a rule that $M$ cells may die due to the absence of long-range signals that are necessary for their survival \cite{Nakamasu} when blue $I^\text{l}$ cells are not present nearby. We adapt this rule as follows:
\begin{align*}  &\frac{\sum_{j=1}^{N_\text{M}}{ \mathds{1}}_{\Omega^{\textbf{M}_i}_{\text{long},\sigma}} (\textbf{M}_i)}{\sum_{i=1}^{N^\text{d}_\text{X}}{ \mathds{1}}_{\Omega^{\textbf{M}_i}_{\text{long},\sigma}}(\textbf{X}^\text{d}_j)} \ge \xi  \hskip0.5cm \text{and  }\hskip0.2cm
\sum_{j=1}^{N_\text{I}^\text{d}} {\mathds{1}}_{B^{\textbf{M}_i}_{90/2,\sigma}} (\textbf{I}_j^\text{l} )< \nu \hskip0.2cm \Longrightarrow \hskip0.2cm \text{death of cell at $\textbf{M}_j$ with probability $p_{\text{death}}$ per day.}
\end{align*}

In the same way, we adapt the model rule for xanthophore form transitions from \cite{volkening2} by replacing the deterministic interaction neighborhoods in the original rules with their stochastic equivalents and reframing the rules in terms of ratios:
\begin{align} &\frac{\sum_{j=1}^{N^\text{l}_\text{I}}{ \mathds{1}}_{B_{75,\sigma}^{\textbf{X}^\text{d}_i}} (\textbf{I}^\text{l}_j)}{a + \sum_{j=1}^{N^\text{d}_\text{I}}{ \mathds{1}}_{B_{90/2,\sigma}^{\textbf{X}^\text{d}_i}} (\textbf{I}_j^\text{d})}  > 1 \hskip0.2cm \Longrightarrow \hskip0.2cm \text{$\textbf{X}^\text{d}_i$ becomes loose}, \label{EqX1} \\
&\frac{\sum_{j=1}^{N^\text{d}_\text{I}}{ \mathds{1}}_{B_{90/2,\sigma}^{\textbf{X}^\text{l}_i}} (\textbf{I}^\text{d}_j) + P_i\sum_{j=1}^{N^\text{d}_\text{X}}{ \mathds{1}}_{B_{75,\sigma}^{\textbf{X}^\text{l}_i}} (\textbf{X}^\text{d}_j)}{ b + \sum_{j=1}^{N^\text{l}_\text{I}}{ \mathds{1}}_{B_{90/2,\sigma}^{\textbf{X}^\text{l}_i}} (\textbf{}\textbf{I}^\text{l}_j )+ \sum_{j=1}^{N_\text{M}}{ \mathds{1}}_{B_{90,\sigma}^{\textbf{X}^\text{l}_i}} (\textbf{M}_j)}  >1 \hskip0.2cm \Longrightarrow \hskip0.2cm \text{$\textbf{X}^\text{l}_i$ becomes dense} \label{EqX2}.
\end{align}

Lastly, we adapt the rules proposed in \cite{volkening2} for iridophore form transitions (between dense and loose) by rescaling by the area of the interaction neighborhoods so that these rules are now in terms of density counts as follows:
\begin{align} 
&\left(\frac{1}{\left|B_{90,\sigma}^{\textbf{I}^\text{l}_i} \right|}\sum_{j=1}^{N_\text{M}}{\mathds{1}}_{B_{90,\sigma}^{\textbf{I}^\text{l}_i}} (\textbf{M}_j)  < \frac{c}{\left|B_{90}^{\textbf{I}^\text{l}_i} \right|} \hskip0.2cm \text{and} \hskip0.2cm \frac{1}{\left|\Omega_{\text{long},\sigma}^{\textbf{I}^\text{l}_i}\right|}\sum_{j=1}^{N^\text{d}_\text{X}}{ \mathds{1}}_{\Omega_{\text{long},\sigma}^{\textbf{I}^\text{l}_i}} (\textbf{X}_j^\text{d})<\frac{d}{\left|\Omega_{\text{long}}^{\textbf{I}^\text{l}_i}\right|}\right) \hskip0.2cm \text{or} \nonumber \\
&\left(\frac{1}{\left|B_{90,\sigma}^{\textbf{I}^\text{l}_i} \right|}\sum_{j=1}^{N_\text{M}}{ \mathds{1}}_{B_{90,\sigma}^{\textbf{I}^\text{l}_i}} (\textbf{M}_j)  < \frac{c}{\left|B_{90}^{\textbf{I}^\text{l}_i} \right|} \hskip0.2cm \text{and}\hskip0.2cm \frac{1}{\left|B_{75,\sigma}^{\textbf{I}^\text{l}_i} \right|}\sum_{j=1}^{N^\text{d}_\text{X}}{ \mathds{1}}_{B_{75,\sigma}^{\textbf{I}^\text{l}_i}} (\textbf{X}_j^\text{d})>\frac{e}{\left|B_{75}^{\textbf{I}^\text{l}_i} \right|}\right) \hskip0.2cm \Longrightarrow \hskip0.2cm \textbf{I}^\text{l}_i \text{ transforms to dense,}\\
&\left(\frac{1}{\left|B_{90,\sigma}^{\textbf{I}^\text{d}_i}\right|}\sum_{j=1}^{N_\text{M}}{ \mathds{1}}_{B_{90,\sigma}^{\textbf{I}^\text{d}_i}} (\textbf{M}_j ) > \frac{f}{\left|B_{90}^{\textbf{I}^\text{d}_i}\right|}\right)\hskip0.2cm \text{or} \nonumber \\ &\left(\frac{1}{\left|\Omega_{\text{long},\sigma}^{\textbf{I}^\text{d}_i}\right|}\sum_{j=1}^{N^\text{d}_\text{X}}{ \mathds{1}}_{\Omega_{\text{long},\sigma}^{\textbf{I}^\text{d}_i}} (\textbf{X}_j^\text{d})>\frac{g}{\left|\Omega_{\text{long}}^{\textbf{I}^\text{d}_i}\right|} \hskip0.2cm \text{and} \hskip0.2cm \frac{1}{\left|B_{75,\sigma}^{\textbf{I}^\text{d}_i}\right|}\sum_{j=1}^{N^\text{d}_\text{X}}{ \mathds{1}}_{B_{75,\sigma}^{\textbf{I}^\text{d}_i}} (\textbf{X}^\text{d}_j )< \frac{h}{\left|B_{75}^{\textbf{I}^\text{d}_i}\right|}\right) \hskip0.2cm \Longrightarrow \hskip0.2cm\textbf{I}^\text{d}_i\text{ transforms to loose},\label{eq:ichange}
\end{align}
where $\left|R\right|$ denotes the area of the interaction neighborhood $R$.

\section*{Supplementary Tables}

\begin{table}[ht!]
    \caption{Summary of the pattern features that we focus on and the corresponding methods that we propose for automatically quantifying these features. We define the application-specific thresholds $T_p^0$, $ T_p^1$, and $T_b^1$ in the main text. We note that $r_b$ is the radius value at which a given topological feature (e.g., connected component or loop) is born and $r_d$ is the radius value at which it disappears.}
    \centering
    \footnotesize{
    \begin{tabular}{lr}
     \textbf{Pattern feature} & \textbf{Quantification method} \\
        \toprule
     Number of spots &  $\beta_0 =  \text{number of dimension $0$ topological generators with persistence } r_d-r_b \geq T_p^0$ \\ 
     Number of stripes &  $\beta_1 =  \text{number of dimension $1$ topological generators with persistence } r_d-r_b \geq T_p^0 \text{ and birth } r_b \leq T_b^1 $  \\ 
       Stripe breaks & Check if $\beta_1 \leq \text{expected number of stripes}$\\ 
       Stripe width &  $r_d-r_b$ of the dimension $1$ persistence points \\ 
       Spot size & Number of agents per single-linkage cluster \\
      Stripe straightness & Arc length distances for single-linkage clusters representing stripes (see \eqref{eq:curviness} in the main text)\\
      Spot roundness & Ratio of PCA eigenvalues for each single-linkage cluster representing a spot (see \eqref{eq:PCA} in the main text) \\ 
      Spot alignment & Nearest-neighbor distances between single-linkage cluster centroids \\ 
      Center radius & Distances from single-linkage cluster centroids to the domain midline \\ 
       Time of stripe formation & Automated boundary search\\ 
       Agent--agent distances & Direct calculations \\
      Agent coefficient of variation (CV)\cite{ParTur256,ParTur130} & Direct calculation using \eqref{eq:CV}  \\ 
      Agent density counts  & Direct calculation of local counts \\ 
       \bottomrule 
    \end{tabular}
    }
    \label{tab:methods_summary}
\end{table}

\begin{table}[h!]
     \centering
    \caption{Quantifying variability and breakdown for wild-type zebrafish patterns as a function of additional stochasticity in cell interactions (also see Fig.~6 in the main text). We base these measurements on $1,000$ simulations (for each value of $\sigma$) of our adapted version of the model \cite{volkening2}, adjusted as described in \textit{Summary of Model Rules Adjusted in our Study of Pattern Variability}. We choose cell-interaction length-scale parameters randomly each day per cell and interaction from a normal distribution (centered at the default length scale) with increasing standard deviation given by $\sigma$ times the default length scale.}
    \begin{tabular}{lrrrrrr} 
   \multicolumn{1}{c}{}&\multicolumn{6}{c}{$\gleftrightarrow{\text{less stochasticity} \hspace{3.5cm} \text{more stochasticity}}$} \\
         \textbf{Noise strength $\sigma$} & \textbf{0.01} & \textbf{0.05} & \textbf{0.1} & \textbf{0.2} & \textbf{0.3} & \textbf{0.5} \\
     \midrule
     Percent of patterns with no breaks & $80.70$\% & $82.70$\% & $80.40$\% & $72.50$\% & $64.90$\% & $24.80$\% \\ 
     Percent of patterns with stripe breaks only   & $3.90$\% & $3.50$\% & $4.40$\% & $5.60$\% & $8.30$\% & $45.90$\% \\ 
      Percent of patterns with interstripe breaks only & $13.70$\% & $11.80$\% & $11.90$\% & $18.60$\% & $20.80$\% & $9.6$0\% \\
      Percent of patterns with stripe \& interstripe breaks & $1.7$0\% & $2.00$\% & $3.30$\% & $3.30$\% & $6.00$\% & $19.70$\%\\
      Mean maximum stripe width ($\mu$m) &  $899.74$ & $892.82$ &  $895.82$ & $911.58$ & $932.44$ & $894.04$ \\
      Standard deviation in maximum stripe width ($\mu$m)  & $27.64$ &  $29.45$ & $ 29.54$ & $31.33$ & $34.80$ & $46.50$ \\ 
      Mean maximum interstripe width ($\mu$m) & $766.02$ &  $780.12$ & $792.54$ &   $820.20$ & $844.60$ &   $903.34$ \\
      Standard deviation in maximum interstripe width ($\mu$m) & $45.88$ & $ 44.82$ & $43.87$ & $ 46.34$ & $56.35$ & $ 83.07$ \\ 
      Mean stripe curviness measure & $7.21$\% & $7.28$\% & $7.50$\% & $8.41$\% & $9.87$\% & $11.59$\%\\
      Standard deviation in stripe curviness & $1.32$\% & $1.24$\% & $1.22$\% & $1.21$\% & $1.36$\% & $4.27$\% \\ 
      Mean time of interstripe X1D \& X1V formation (dpf) & $41.14$  & $40.21$ & $40.40$ & $40.64$ & $41.11$ & $42.29$ \\
     Standard deviation in X1D \& X1V formation (dpf) & $1.26$ & $1.24$ & $1.26$ &  $1.29$ &  $1.31$ & $1.55$ \\ 
      \bottomrule
    \end{tabular}
    \label{tab:WT_pcpd}
\end{table}

\begin{table}[h!]
    \centering
        \caption{Quantifying pattern variability and breakdown for the \emph{nacre} mutant as a function of additional stochasticity in cell interactions (also see Fig.~6 in the main text). We base these measurements on $1,000$ simulations (for each value of $\sigma$) of our adapted version of the model \cite{volkening2}, adjusted as described in \textit{Summary of Model Rules Adjusted in our Study of Pattern Variability}. We choose cell-interaction length-scale parameters each day randomly per cell and interaction from a normal distribution (centered at the default length scale) with increasing standard deviation given by $\sigma$ times the default length scale.}
    \begin{tabular}{lrrrrrr}
       \multicolumn{1}{c}{}&\multicolumn{6}{c}{$\gleftrightarrow{\text{less stochasticity} \hspace{3.5cm} \text{more stochasticity}}$} \\
          \textbf{Noise strength $\sigma$} & \textbf{0.01} & \textbf{0.05} & \textbf{0.1} & \textbf{0.2} & \textbf{0.3} & \textbf{0.5} \\
     \midrule
     Percent of patterns with normal spots & $99.40$\% & $98.80$\% & $97.50$\% & $45.50$\% & $0.00$\% & $0.00$\%  \\
     Percent of patterns with small spots & $0.10$\% & $0.00$\% & $1.40$\% & $54.30$\% & $27.00$\% & $0.00$\% \\ 
     Percent of patterns with many spots & $0.50$\% & $1.20$\% & $1.00$\% & $0.00$\% & $0.00$\% & $0.00$\% \\ 
     Percent of patterns with small spots and many spots & $0.00$\% & $0.00$\% & $0.10$\% & $0.20$\% & $73.0$\% & $100.0$\% \\ 
     Mean spot roundness score &  $6.45$ &  $6.07$ &  $5.81$ & $4.81$ & $4.22$ & $10.59$ \\ 
     Standard deviation in spot roundness score &  $6.17$ &  $6.03$ &  $5.63$ & $4.31$ & $3.26$ & $20.23$ \\
     Mean variance in spot spacing ($\mu$m) &  $337.17$ & $339.36$ & $344.00$ & $375.56$ & $388.81$ & $148.18$ \\ 
     Standard deviation in variance in spot spacing ($\mu$m)  & $67.07$ & $72.56$ & $75.74$ &  $108.11$ &  $143.95$ & 64.99 \\
     Mean X0 interstripe width ($\mu$m) & $1044.92$ & $1064.99$ & $1073.66$ &  $1338.23$ &  $2860.02$ &  $2001.47$ \\ 
     Standard deviation in X0 interstripe width ($\mu$m) & $416.42$ &  $448.59$ &  $441.43$ &  $672.62$ &   $973.30$ &$1171.27$ \\ 
     \bottomrule
\end{tabular}
    \label{tab:nacre_pcpd}
\end{table}

\begin{table}[h!]
    \centering
        \caption{Quantifying pattern variability and breakdown for the \emph{pfeffer} mutant as a function of additional stochasticity in cell interactions (also see Fig.~6 in the main text). We base these measurements on $1,000$ simulations (for each value of $\sigma$) of our adapted version of the model \cite{volkening2}, adjusted as described in \textit{Summary of Model Rules Adjusted in our Study of Pattern Variability}. We choose cell-interaction length-scale parameters each day randomly per cell and interaction from a normal distribution (centered at the default length scale) with increasing standard deviation given by $\sigma$ times the default length scale.}
    \begin{tabular}{lrrrrrr}
       \multicolumn{1}{c}{}&\multicolumn{6}{c}{$\gleftrightarrow{\text{less stochasticity} \hspace{3.5cm} \text{more stochasticity}}$} \\
      \textbf{Noise strength $\sigma$} & $\textbf{0.01}$ & \textbf{0.05} & \textbf{0.1} & \textbf{0.2} & \textbf{0.3} & \textbf{0.5} \\
     \midrule
     Percent of patterns with normal spots & $82.30$\% & $88.40$\% &  $83.50$\% & $15.70$\% &$ 0.10$\% & $0.00$\% \\ 
     Percent of patterns with small spots & $17.70$\% & $11.40$\% & $16.30$\% & $75.80$\% & $16.80$\% &$0.70$\%  \\ 
     Percent of patterns with few spots & $0.00$\% & $0.10$\% & $0.20$\% & $0.0$0\% & $0.00$\% & $0.00$\% \\
     Percent of patterns with small spots and many spots & $0.00$\% & $0.10$\% & $0.00$\% & $8.50$\% & $83.10$\% & $99.30$\% \\ 
      Mean spot roundness score & $2.32$ & $2.30$ & $2.29$ & $2.42$ & $2.76$ &  $4.97$ \\ 
     Standard deviation in spot roundness score  & $0.67$ & $0.62$ &  $0.64$ & $0.76$ & $1.14$ & $7.68$ \\ 
     Mean variance in spot spacing ($\mu$m) & $330.29$ & $333.95$ & $337.69$ &  $325.61$ & $290.93$ & $286.42$ \\
     Standard deviation in variance in spot spacing ($\mu$m) & $30.23$ & $30.18$ & $31.14$ &  $37.26$ &  $50.14$ &  $101.07$ \\
     Mean X0 interstripe width ($\mu$m) & $920.98$ &  $919.72$ & $936.03$ & $1000.12$ & $1132.60$ & $2304.91$ \\
      Standard deviation in X0 interstripe width ($\mu$m)  & $155.33$ & $140.40$ & $150.40$ &  $187.53$ & $386.15$ & $1040.17$\\ 
      \bottomrule 
\end{tabular}
    \label{tab:pfef_pcpd}
\end{table}

\begin{table}[h!]
    \centering
        \caption{Quantifying pattern variability and breakdown for the \emph{shady} mutant as a function of additional stochasticity in cell interactions (also see Fig.~6 in the main text). We base these measurements on $1,000$ simulations (for each value of $\sigma$) of our adapted version of the model \cite{volkening2}, adjusted as described in \textit{Summary of Model Rules Adjusted in our Study of Pattern Variability}. We choose cell-interaction length-scale parameters each day randomly per cell and interaction from a normal distribution (centered at the default length scale) with increasing standard deviation given by $\sigma$ times the default length scale.}
      \begin{tabular}{lrrrrrr}
       \multicolumn{1}{c}{}&\multicolumn{6}{c}{$\gleftrightarrow{\text{less stochasticity} \hspace{3.5cm} \text{more stochasticity}}$} \\
     \textbf{Noise strength $\sigma$} & \textbf{0.01} & \textbf{0.05} & \textbf{0.1} & \textbf{0.2} & \textbf{0.3} & \textbf{0.5} \\
     \midrule
     Percent of patterns with normal spots & $100.00$\% & $99.30$\% & $97.30$\% & $47.80$\% & $1.40$\% & $0.00$\% \\
     Percent of patterns with small spots & $0.00$\% & $0.20$\% & $0.00$\% & $0.00$\% & $0.30$\% & $0.00$\% \\ 
     Percent of patterns with few spots & $0.00$\%  & $0.40$\% & $2.70$\% & $52.20$\% & $86.00$\% & $12.60$\% \\ 
     Percent of patterns with no spots & $0.00$\% & $0.00$\% & $0.00$\% & $0.00$\% & $0.10$\% & $70.70$\% \\ 
     Percent of patterns with small spots and few spots & $0.00$\% & $0.00$\% & $0.00$\% & $0.00$\% & $11.90$\% & $16.60$\% \\ 
     Percent of patterns with big spots and few spots & $0.00$\% & $0.10$\% & $0.00$\% & $0.00$\% & $0.30$\% & $0.10$\% \\
      Mean roundness score & $2.85$ & $2.68$ &  $2.43$ &  $2.13$ &  $2.55$ & $2.53$\\
      Standard deviation in spot roundness score & $1.08$ & $0.98$ & $0.86$ & $0.86$ &  $2.14$ & $1.22$ \\ 
     Mean variance in spot spacing ($\mu$m) & $455.06$ &  $462.97$ & $472.76$ & $537.76$ & $789.41$ & N/A \\
     Standard deviation in variance in spot spacing ($\mu$m) &   $50.02$ &  $53.29$ &  $56.17$ & $96.30$ & $279.46$ & N/A \\
     Mean X0 interstripe width ($\mu$m) & $1239.53$ & $1239.42$ & $1247.12$ & $1366.71$ &  $2044.69$ &$4278.00$ \\
     Standard deviation in X0 interstripe width ($\mu$m) & $188.39$ & $191.90$ &  $192.84$ &  $262.18$ & $820.31$ & $0.00$ \\
     \bottomrule
\end{tabular}
    \label{tab:shady_pcpd}
\end{table}

\clearpage

\bibliography{zebrafishBib}

\begin{thebibliography}{10}

\bibitem{ChoCancer}
H.~Cho and D.~Levy.
\newblock Modeling the dynamics of heterogeneity of solid tumors in response to
  chemotherapy.
\newblock {\em Bull. Math. Biol.}, 79(12):2986--3012, 2017.

\bibitem{WoundReview}
B.~D. Cumming, D.~L. McElwain, and Z.~Upton.
\newblock A mathematical model of wound healing and subsequent scarring.
\newblock {\em J. Royal Soc. Interface}, 7(42):19--34, 2010.

\bibitem{WangHair}
Qixuan Wang, Ji~Won Oh, Hye-Lim Lee, Tao Dhar, Anukriti an d~Peng, Raul Ramos,
  Christian~Fernando Guerrero-Juarez, Xiaojie Wang, Ran Zhao, Xiaoling Cao,
  Jonathan Le, Meli sa~A Fuentes, Shelby~C Jocoy, Antoni~R Rossi, Brian Vu, Kim
  Pham, Xiaoyang Wang, Nanda~Maya Mali, Jung~Min Park, June-Hyug Choi, Hyunsu
  Lee, Julien M~D Legrand, Eve Kandyba, Jung~Chul Kim, Moonkyu Kim, John Foley,
  Zhengquan Yu, Bogi Kobielak, Krzysztof a nd~Andersen, Kiarash Khosrotehrani,
  Qing Nie, and Maksim~V Plikus.
\newblock A multi-scale model for hair follicles reveals heterogeneous domains
  driving rapid spatiotemporal hair growth patterning.
\newblock {\em Elife}, 6:e22772, 2017.

\bibitem{GloverHair}
James~D. Glover, Kirsty~L. Wells, Franziska Matth\"{a}us, Kevin~J. Painter,
  William Ho, Jon Riddell, Jeanette~A. Johansson, Matthew~J. Ford, Colin A.~B.
  Jahoda, Vaclav Klika, Richard~L. Mort, and Denis~J. Headon.
\newblock Hierarchical patterning modes orchestrate hair follicle
  morphogenesis.
\newblock {\em PLOS Biol.}, 15(7):1--31, 2017.

\bibitem{Jan}
A.~P. Singh and C.~N{\"u}sslein-Volhard.
\newblock Zebrafish stripes as a model for vertebrate colour pattern formation.
\newblock {\em Curr. Biol.}, 25(2):R81--R92, 2015.

\bibitem{kondoTuringQuestion}
M.~Watanabe and S.~Kondo.
\newblock Is pigment patterning in fish skin determined by the {T}uring
  mechanism?
\newblock {\em Trends Genet.}, 31(2):88--96, 2015.

\bibitem{McMen2016}
S.~K. McMenamin, M.~N. Chandless, and D.~M. Parichy.
\newblock Working with zebrafish at postembryonic stages.
\newblock {\em Methods Cell Biol.}, 134:587--607, 2016.

\bibitem{irion2016chapter}
U.~Irion, A.~P. Singh, and C.~N{\"u}sslein-Volhard.
\newblock The developmental genetics of vertebrate color pattern formation:
  lessons from zebrafish.
\newblock {\em Curr. Top. Dev. Biol.}, 117:141--169, 2016.

\bibitem{Yamaguchi}
M.~Yamaguchi, E.~Yoshimoto, and S.~Kondo.
\newblock Pattern regulation in the stripe of zebrafish suggests an underlying
  dynamic and autonomous mechanism.
\newblock {\em Pro. Natl. Acad. Sci. U.S.A.}, 104(12):4790--4793, 2007.

\bibitem{Maderspacher2003}
F.~Maderspacher and C.~N{\"u}sslein-Volhard.
\newblock Formation of the adult pigment pattern in zebrafish requires
  \emph{leopard} and \emph{obelix} dependent cell interactions.
\newblock {\em Development}, 130(15):3447--3457, 2003.

\bibitem{ParTur130}
D.~M. Parichy and J.~M. Turner.
\newblock Temporal and cellular requirements for fms signaling during zebrafish
  adult pigment pattern development.
\newblock {\em Development}, 130(5):817--833, 2003.

\bibitem{Frohnhofer}
H.~G. Frohnh{\"o}fer, J.~Krauss, H.~M. Maischein, and C.~N{\"u}sslein-Volhard.
\newblock Iridophores and their interactions with other chromatophores are
  required for stripe formation in zebrafish.
\newblock {\em Development}, 140(14):2997--3007, 2013.

\bibitem{Lister}
J.~A. Lister, C.~P. Robertson, T.~Lepage, S.~L. Johnson, and D.~W. Raible.
\newblock \emph{nacre} encodes a zebrafish microphthalmia-related protein that
  regulates neural-crest-derived pigment cell fate.
\newblock {\em Development}, 126(17):3757--3767, 1999.

\bibitem{PatDev127}
D.~M. Parichy, D.~G. Ransom, B.~Paw, L.~I. Zon, and S.~L. Johnson.
\newblock An orthologue of the kit-related gene fms is required for development
  of neural crest-derived xanthophores and a subpopulation of adult melanocytes
  in the zebrafish, \emph{{D}anio rerio}.
\newblock {\em Development}, 127(14):3031--3044, 2000.

\bibitem{Lopes}
Susana~S. Lopes, Xueyan Yang, Jeanette M\"{u}ller, Thomas~J. Carney, Anthony~R.
  McAdow, Gerd-J\"{o}rg Rauch, Arie~S. Jacoby, Laurence~D. Hurst, Mariana
  Delfino-Mach\'{i}­n, Pascal Haffter, Robert Geisler, Stephen~L. Johnson,
  Andrew Ward, and Robert~N. Kelsh.
\newblock Leukocyte tyrosine kinase functions in pigment cell development.
\newblock {\em PLOS Genet.}, 4(3):1--13, 03 2008.

\bibitem{Nakamasu}
A.~Nakamasu, G.~Takahashi, A.~Kanbe, and S.~Kondo.
\newblock Interactions between zebrafish pigment cells responsible for the
  generation of {T}uring patterns.
\newblock {\em Pro. Natl. Acad. Sci. U.S.A.}, 106(21):8429--8434, 2009.

\bibitem{Painter}
K.~J. Painter, J.~M. Bloomfield, J.~A. Sherratt, and A.~Gerisch.
\newblock A nonlocal model for contact attraction and repulsion in
  heterogeneous cell populations.
\newblock {\em Bull. Math. Biol.}, 77(6):1132--1165, 2015.

\bibitem{Bullara}
D.~Bullara and Y.~De~Decker.
\newblock Pigment cell movement is not required for generation of {T}uring
  patterns in zebrafish skin.
\newblock {\em Nat. Commun.}, 6, 2015.

\bibitem{MorDeutsch}
J.~Moreira and A.~Deutsch.
\newblock Pigment pattern formation in zebrafish during late larval stages: a
  model based on local interactions.
\newblock {\em Dev. Dyn.}, 232(1):33--42, 2005.

\bibitem{volkening}
A.~Volkening and B.~Sandstede.
\newblock Modelling stripe formation in zebrafish: an agent-based approach.
\newblock {\em J. Royal Soc. Interface}, 12, 2015.

\bibitem{volkening2}
A.~Volkening and B.~Sandstede.
\newblock Iridophores as a source of robustness in zebrafish stripes and
  variability in \emph{Danio} patterns.
\newblock {\em Nat. Commun.}, 9(3231), 2018.

\bibitem{Shinbrot}
C.~E. Caicedo-Carvajal and T.~Shinbrot.
\newblock \emph{In silico} zebrafish pattern formation.
\newblock {\em Dev. Biol.}, 315(2):397--403, 2008.

\bibitem{Lee2018}
D.~E. Lee, D.~R. Cavener, and M.~L. Bond.
\newblock Seeing spots: quantifying mother-offspring similarity and assessing
  fitness consequences of coat pattern traits in a wild population of giraffes
  (\emph{{G}iraffa camelopardalis}).
\newblock {\em PeerJ}, 2018.

\bibitem{imageJ}
C.~A. Schneider, W.~S. Rasband, and K.~W. Eliceiri.
\newblock {NIH} image to {ImageJ}: 25 years of image analysis.
\newblock {\em Nat. Methods}, 9:671--675, 2012.

\bibitem{Miyazawa2010}
S.~Miyazawa, M.~Okamoto, and S.~Kondo.
\newblock Blending of animal colour patterns by hybridization.
\newblock {\em Nat. Commun.}, 1(66), 2010.

\bibitem{Djurdjevic2019}
I.~Djurdjevic, T.~Furmanek, S.~Miyazawa, and S.~S. Bajec.
\newblock Comparative transcriptome analysis of trout skin pigment cells.
\newblock {\em BMC Genom.}, 20(359), 2019.

\bibitem{delta}
Hiroki Hamada, Masakatsu Watanabe, Hiu~Eunice Lau, Tomoki Nishida, Toshiaki
  Hasegawa, David~M Parichy, and Shigeru Kondo.
\newblock Involvement of delta/notch signaling in zebrafish adult pigment
  stripe patterning.
\newblock {\em Development}, 141(2):318--324, 2014.

\bibitem{Topaz2015}
C.~M. Topaz, L.~Ziegelmeier, and T.~Halverson.
\newblock Topological data analysis of biological aggregation models.
\newblock {\em PLOS One}, 10(5):1--26, 05 2015.

\bibitem{hirata2005pigment}
M.~Hirata, K.~Nakamura, and S.~Kondo.
\newblock Pigment cell distributions in different tissues of the zebrafish,
  with special reference to the striped pigment pattern.
\newblock {\em Dev. Dyn.}, 234(2):293--300, 2005.

\bibitem{McMen2014}
Sarah~K McMenamin, Emily~J Bain, Anna~E McCann, Larissa~B Patterson, Dae~Seok
  Eom, Zachary~P Waller, James~C Hamill, Julie~A Kuhlman, Judith~S Eisen, and
  David~M Parichy.
\newblock Thyroid hormone--dependent adult pigment cell lineage and pattern in
  zebrafish.
\newblock {\em Science}, 345(6202):1358--1361, 2014.

\bibitem{Mahalwar}
P.~Mahalwar, B.~Walderich, A.~P. Singh, and C.~N{\"u}sslein-Volhard.
\newblock Local reorganization of xanthophores fine-tunes and colors the
  striped pattern of zebrafish.
\newblock {\em Science}, 345(6202):1362--1364, 2014.

\bibitem{Patterson2013}
L.~B. Patterson and D.~M. Parichy.
\newblock Interactions with iridophores and the tissue environment required for
  patterning melanophores and xanthophores during zebrafish adult pigment
  stripe formation.
\newblock {\em PLOS Genet.}, 9(5), 2013.

\bibitem{2016heterotypic}
P.~Mahalwar, A.~P. Singh, A.~Fadeev, C.~N{\"u}sslein-Volhard, and U.~Irion.
\newblock Heterotypic interactions regulate cell shape and density during color
  pattern formation in zebrafish.
\newblock {\em Biol. Open}, 5(11):1680--1690, 2016.

\bibitem{PatNcomm}
L.~B. Patterson, E.~J. Bain, and D.~M. Parichy.
\newblock Pigment cell interactions and differential xanthophore recruitment
  underlying zebrafish stripe reiteration and \emph{{D}anio} pattern evolution.
\newblock {\em Nat. Commun.}, 5, 2014.

\bibitem{Inaba}
M.~Inaba, H.~Yamanaka, and S.~Kondo.
\newblock Pigment pattern formation by contact-dependent depolarization.
\newblock {\em Science}, 335(6069):677--677, 2012.

\bibitem{eom2015long}
D.~S. Eom, E.~J. Bain, L.~B. Patterson, M.~E. Grout, and D.~M. Parichy.
\newblock Long-distance communication by specialized cellular projections
  during pigment pattern development and evolution.
\newblock {\em Elife}, 4:e12401, 2015.

\bibitem{fadeev2016}
A.~Fadeev, J.~Krauss, A.~P. Singh, and C.~N{\"u}sslein-Volhard.
\newblock Zebrafish leucocyte tyrosine kinase controls iridophore
  establishment, proliferation and survival.
\newblock {\em Pigment Cell Melanoma Res.}, 29(3):284--296, 2016.

\bibitem{Parichy}
D.~M. Parichy, M.~R. Elizondo, M.~G. Mills, T.~N. Gordon, and R.~E. Engeszer.
\newblock Normal table of postembryonic zebrafish development: staging by
  externally visible anatomy of the living fish.
\newblock {\em Dev. Dyn.}, 238(12):2975--3015, 2009.

\bibitem{Singh}
A.~P. Singh, U.~Schach, and C.~N{\"u}sslein-Volhard.
\newblock Proliferation, dispersal and patterned aggregation of iridophores in
  the skin prefigure striped colouration of zebrafish.
\newblock {\em Nat. Cell Biol.}, 16(6):604--611, 2014.

\bibitem{TakahashiMelDisperse}
G.~Takahashi and S.~Kondo.
\newblock Melanophores in the stripes of adult zebrafish do not have the nature
  to gather, but disperse when they have the space to move.
\newblock {\em Pigment Cell Melanoma Res.}, 21(6):677--686, 2008.

\bibitem{Yamanaka2014}
H.~Yamanaka and S.~Kondo.
\newblock In vitro analysis suggests that difference in cell movement during
  direct interaction can generate various pigment patterns in vivo.
\newblock {\em Pro. Natl. Acad. Sci. U.S.A.}, 111(5):1867--1872, 2014.

\bibitem{ParTur256}
D.~M. Parichy and J.~M Turner.
\newblock Zebrafish \emph{puma} mutant decouples pigment pattern and somatic
  metamorphosis.
\newblock {\em Dev. Biol.}, 256(2):242--257, 2003.

\bibitem{walderich2016homotypic}
B.~Walderich, A.~P. Singh, P.~Mahalwar, and C.~N{\"u}sslein-Volhard.
\newblock Homotypic cell competition regulates proliferation and tiling of
  zebrafish pigment cells during colour pattern formation.
\newblock {\em Nat. Commun.}, 7, 2016.

\bibitem{Carlsson2009}
G.~Carlsson.
\newblock Topology and data.
\newblock {\em Bull. Am. Math. Soc.}, 46(2):255--308, 2009.

\bibitem{chazal}
F.~Chazal, V.~de~Silva, M.~Glisse, and S.~Oudot.
\newblock {\em The Structure and Stability of Persistence Modules}.
\newblock Springer International Publishing, 1.0 edition, 2016.

\bibitem{Edelsbrunner10}
H.~Edelsbrunner and J.~L. Harer.
\newblock {\em Computational Topology, An Introduction}.
\newblock American Mathematical Society, 2010.

\bibitem{Ghrist2014}
R.~Ghrist.
\newblock {\em Elementary Applied Topology}.
\newblock Createspace, 1.0 edition, 2014.

\bibitem{Zomorodian2009}
A.~Zomorodian.
\newblock {\em Topology for Computing}.
\newblock Cambridge University Press, 2009.

\bibitem{Bishop2006}
C.~M. Bishop.
\newblock {\em Pattern Recognition and Machine Learning}.
\newblock Springer {S}cience$+${B}usiness, New York, NY, 2006.

\bibitem{HastieBook}
T.~Hastie, R.~Tibshirani, and J.~Friedman.
\newblock {\em The Elements of Statistical Learning}.
\newblock Springer, 2009.

\bibitem{genome}
Kerstin Howe, Matthew~D Clark, Carlos~F Torroja, James Torrance, Camille
  Berthelot, Matthieu Muffato, John~E Collins, Sean Humphray, Karen McLaren,
  Lucy Matthews, et~al.
\newblock The zebrafish reference genome sequence and its relationship to the
  human genome.
\newblock {\em Nature}, 496(7446):498--503, 2013.

\bibitem{Budi}
E.~H. Budi, L.~B. Patterson, and D.~M. Parichy.
\newblock Post-embryonic nerve-associated precursors to adult pigment cells:
  genetic requirements and dynamics of morphogenesis and differentiation.
\newblock {\em PLoS Genet.}, 7(5):e1002044, 2011.

\bibitem{Dooley}
C.~M. Dooley, A.~Mongera, B.~Walderich, and C.~N{\"u}sslein-Volhard.
\newblock On the embryonic origin of adult melanophores: the role of erbb and
  kit signalling in establishing melanophore stem cells in zebrafish.
\newblock {\em Development}, 140(5):1003--1013, 2013.

\bibitem{Ripser}
U.~Bauer.
\newblock Ripser: {E}fficient computation of vietoris--rips persistence
  barcodes, August 2019.
\newblock 1908.02518.

\bibitem{Curto2017}
C.~Curto.
\newblock What can topology tell us about the neural code?
\newblock {\em Bull. Am. Math. Soc.}, 54(1):63--78, 2017.

\bibitem{Giusti13455}
C.~Giusti, E.~Pastalkova, C.~Curto, and V.~Itskov.
\newblock Clique topology reveals intrinsic geometric structure in neural
  correlations.
\newblock {\em Proc. Natl. Acad. Sci. U.S.A.}, 112(44):13455--13460, 2015.

\bibitem{Sizemore2018}
A.~E. Sizemore, C.~Giusti, A.~Kahn, J.~M. Vettel, R.~F. Betzel, and D.~S.
  Bassett.
\newblock Cliques and cavities in the human connectome.
\newblock {\em J. Comput. Neurosci.}, 44(1):115--145, 2018.

\bibitem{Chan2013}
J.~M. Chan, G.~Carlsson, and R.~Rabadan.

\bibitem{camera2016}
P.~G. Camara, D.~I.S. Rosenbloom, K.~J. Emmett, A.~J. Levine, and R.~Rabadan.
\newblock Topological data analysis generates high-resolution, genome-wide maps
  of human recombination.
\newblock {\em Cell Systems}, 3(1):83 -- 94, 2016.

\bibitem{Humphreys2019}
D.~P. Humphreys, M.~R. McGuirl, M.~Miyagi, and A.~J. Blumberg.
\newblock Fast estimation of recombination rates using topological data
  analysis.
\newblock {\em Genetics}, 211:1--14, 2019.

\bibitem{Munch2012}
E.~Munch, M.~Shapiro, and J.~Harer.
\newblock Failure filtrations for fenced sensor networks.
\newblock {\em Int. J. Robotics Res.}, 31:1044--1056, 2012.

\bibitem{Adams2015}
H.~Adams and G.~Carlsson.
\newblock Evasion paths in mobile sensor networks.
\newblock {\em Int. J. Robotics Res.}, 34:90--104, 2015.

\bibitem{Hatcher:478079}
A.~Hatcher.
\newblock {\em {Algebraic topology}}.
\newblock Cambridge University Press, Cambridge, 2000.

\end{thebibliography}
\bibliographystyle{unsrt}

\end{document}